# Autocatalytic Models for the Origin of Biological Homochirality

Donna G. Blackmond
Department of Chemistry
Scripps Research
La Jolla, CA 92037 USA

Table of Contents

## INTRODUCTION

The single-handedness of the molecular building blocks of biological polymers – D-sugars and L-amino acids – is considered a signature of life. In the absence of a chiral directing force, an abiotic process necessarily yields an equal mixture of left- and right-handed molecules. The question of how biological homochirality could have emerged from a presumably racemic prebiotic soup has intrigued scientists since Pasteur's deliberate separation of the homochiral crystals formed from the di-salt of tartaric acid in 1848.[1] Theoretical approaches[2] to this question preceded experimental investigations until the last decade of the 20th century, when several striking findings demonstrated ways in which enantioenrichment could be achieved through physical processes[3] or chemical reactions.[4] These exciting developments spurred extensive further experimental work so that, by end of the first decade of the 21st century, the variety of reports in the literature prompted the observation that we have become "spoilt for choice"[5] in approaches to rationalize the emergence of homochirality. Some of the earliest and most compelling theoretical proposals for asymmetric amplification of an initial small imbalance of enantiomers invoked autocatalytic reaction systems. The purpose of this review is to summarize studies aimed at understanding how autocatalytic systems may lead to the emergence of homochirality.

## THE FRANK AUTOCATALYTIC MODEL

The classic 1953 paper by Frank presented a mathematical mode of autocatalysis that provides a "simple and sufficient life model" leading ultimately to homochirality over many autocatalytic cycles, as illustrated in Scheme 1. The two key features of this theoretical model are: 1) each enantiomer of a molecule is able to self-replicate and, importantly, 2) each enantiomer is able to suppress, or partially suppress, the replication of its mirror image. This second aspect is accomplished by what Frank termed “mutual antagonism”, which is key to the amplification of enantiomeric excess (*ee*) over many cycles. Frank’s paper ended with the rather understated observation that an experimental demonstration of his mathematical model “may not be impossible.” As later noted by Wynberg, this simple statement served as a call to arms to “every red-blooded synthetic organic chemist.”[6] Here we review the chronology of attempts to demonstrate the Frank autocatalytic model for the emergence of homochirality and to assess the prebiotic plausibility of such systems.

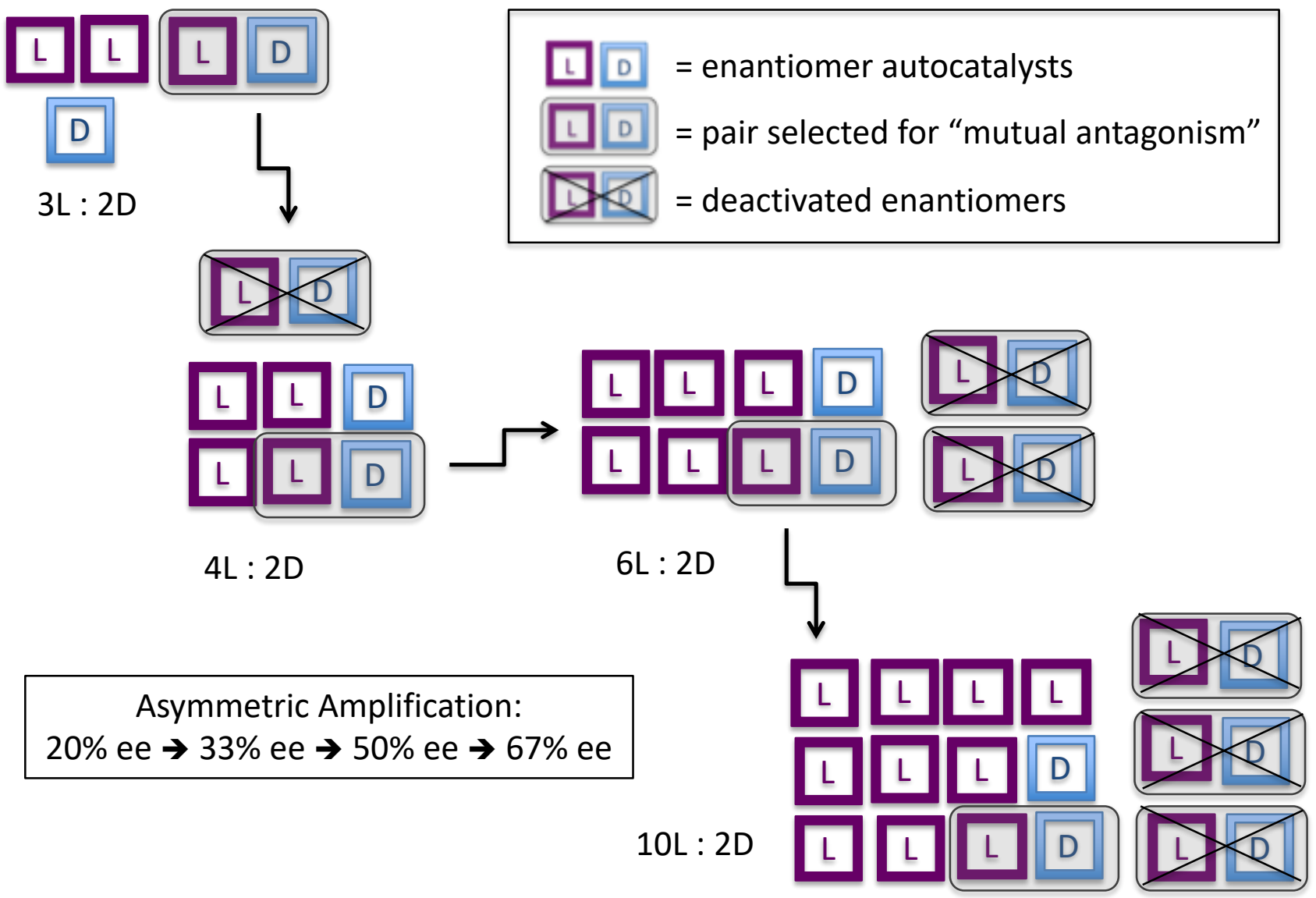

**Scheme 1.** Frank’s model for asymmetric amplification via autocatalytic self-replication and mutual antagonism.

## EARLY EFFORTS TO MEET FRANK’S CHALLENGE

After the appearance of Frank’s 1953 paper, synthetic organic chemists began to speculate about particular chemical systems that might exhibit the features delineated by the Frank model. Initially, most work focused on point 1, autocatalysis, rather than point 2, an inhibition mechanism. Sigmoidal product concentration profiles have often been presented as evidence of autocatalysis. However, sigmoidal kinetic behavior can arise from other causes,[7] which, as shown in Scheme 2, commonly include catalyst activation[8] and product-induced rate acceleration.[9]

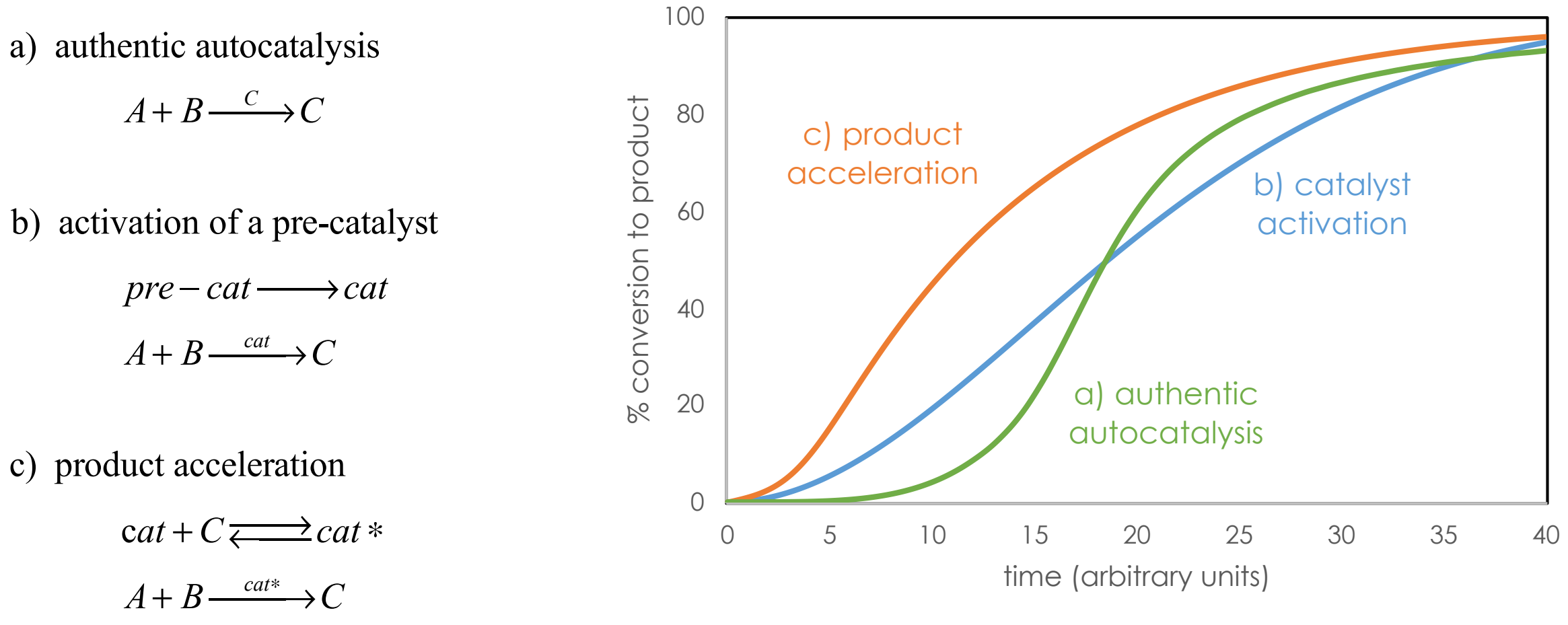

**Scheme 2.** Reaction networks exhibiting sigmoidal profiles. a) authentic autocatalysis, where product *C* catalyzes its own formation; b) catalyst activation, where a pre-catalyst *pre-cat* reacts to form the active catalyst *cat* over the course of the reaction; c) product acceleration, where reaction product *C* binds to the catalyst *cat* to form a more efficient catalyst *cat**. Although all three reaction networks may exhibit sigmoidal kinetic profiles, only an authentic autocatalytic reaction system has the potential to lead to homochirality from a small initial imbalance.

Cases where a catalyst becomes more active and potentially more selective, due either to modifications of a precatalyst or by interaction with the reaction product, do not obey the Frank model. In these reactions, because the total active catalyst concentration is necessarily limited by that of the catalyst originally employed, the asymptotic approach to homochirality predicted by the Frank model cannot occur. Observation of sigmoidal behavior is insufficient to support the proposal of an authentic autocatalytic reaction in the sense of the Frank model. By contrast, in an authentic, truly self-replicating autocatalytic reaction that is continually fed with reactants, catalyst concentration may increase indefinitely, allowing for both an accelerating rate and a continually improving product selectivity that asymptotically approaches homochirality. Thus, one key to the search for a chemical system from which homochirality can emerge is a decisive distinction between these other kinetic behaviors, which may be classified as *autoinductive* catalysis and will not lead to homochirality, and true *self-replicative* autocatalysis, which potentially may lead to homochirality.

A number of early attempts aimed at uncovering autocatalytic reaction systems with potentially prebiotically relevant chemistry have been reported. Breslow's[10] classic finding of autocatalytic behavior in the formose reaction revealed that the achiral product glycolaldehyde catalyzes its own formation (Scheme 3). Glycoaldehdye is achiral, however, and hence its autocatalytic production in the formose reaction cannot lead to homochirality. Prebiotically relevant chiral molecules such as glyceraldehyde are produced in further reactions in this network. Modest enantioselectivity in glyceraldehyde has been observed when the formose reaction is carried out in the presence of chiral amino acids and their derivatives as asymmetric catalysts.[11] However, these catalytic reactions have not been shown to be truly self-replicative and therefore will not result in the emergence of homochirality.

Wynberg[6] suggested that the classic Betti reaction,[12] addition of a carbon nucleophile to an imine, which had been reported to exhibit a temporally increasing reaction rate, might be a candidate for asymmetric autocatalysis. The reaction successfully produces an optically active amine product when carried out in the presence of a chiral amine catalyst (brucine, an optically pure alkaloid), but the reaction product itself was unable to serve as a catalyst in its own formation (Scheme 4).

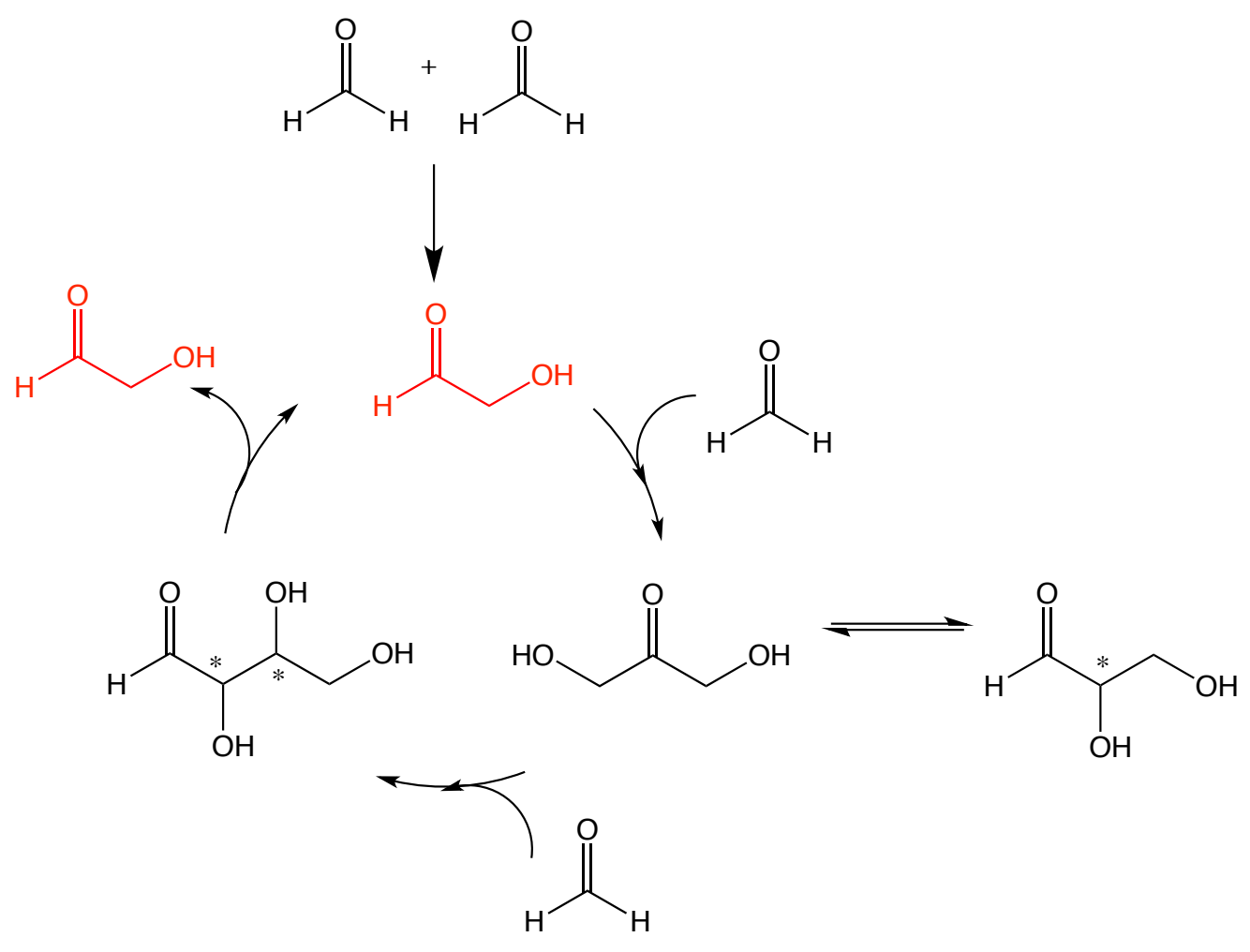

**Scheme 3.** Mechanism of the formose reaction revealing the role of glycolaldehyde (red) in catalyzing it own formation. Chiral $C_3$ and $C_4$ molecules are produced in this reaction, but the autocatalyst is achiral.

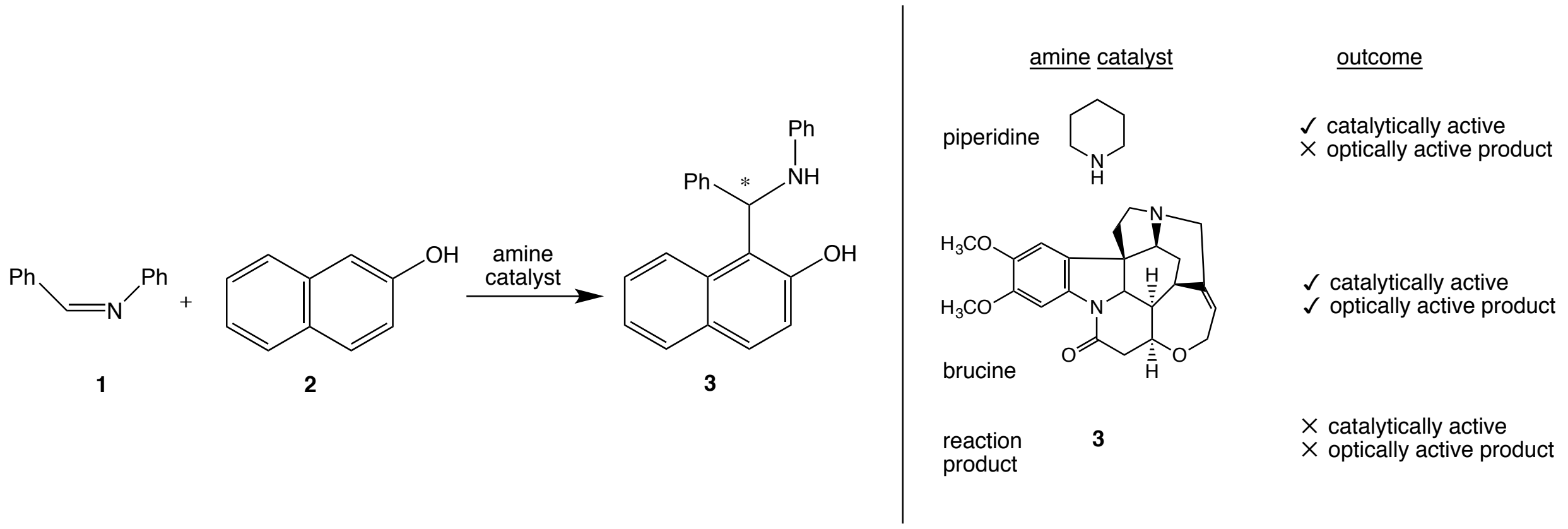

**Scheme 4.** Wynberg's[6] early attempt at an enantioselective, autocatalytic Betti[12] reaction. The reaction product does not serve as a catalyst in its own formation.

Commeyras[13] reported sigmoidal kinetic behavior in the synthesis of amino acids via hydration of aminonitriles. Kinetic studies revealed that carbonyl compounds (acetaldehyde or acetone) formed from partial decomposition of the aminonitrile can act to catalyze the hydration of the remaining aminonitrile (Scheme 5). This reaction presents a product-induced mechanism but does not correspond to a self-replicative autocatalytic model and does not directly offer a mechanism for enantioenrichment.

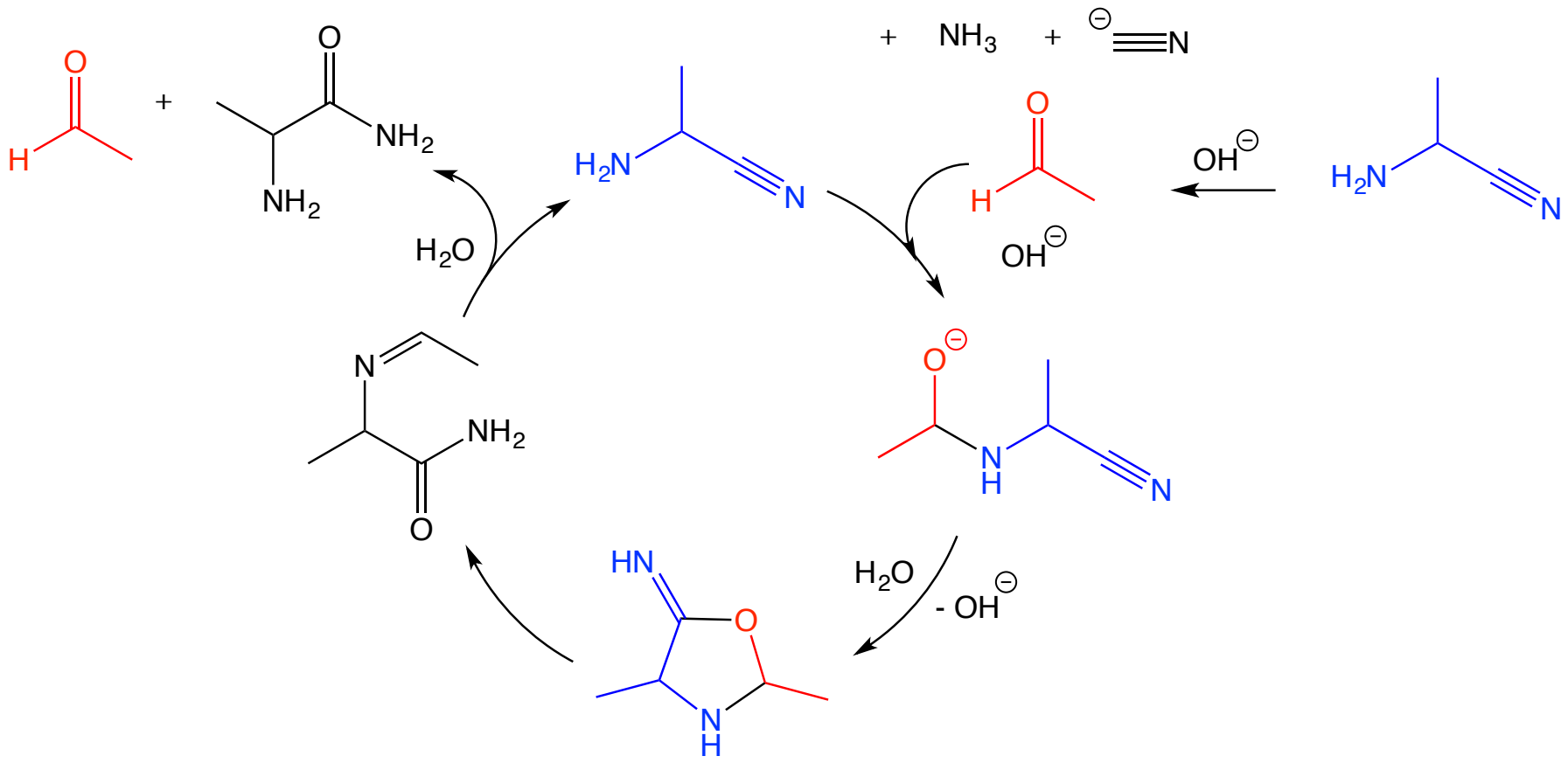

**Scheme 5.** Commeyras and coworkers'[13] autoinductive mechanism for aminonitrile hydration assisted by acetaldehyde formed from partial decomposition of the starting material.

## THE CHALLENGE IS MET: GENESIS OF THE SOAI REACTION

All of the examples discussed above describe reactions that exhibit sigmoidal kinetic profiles, but none meets the criteria set by the Frank model for an asymmetric autocatalytic reaction leading to the emergence of homochirality. In further pursuit of Frank's challenge, a number of groups turned to reactions involving alkyl metal reagents. These molecules can form complex higher-order chiral structures that may introduce nonlinear effects into the reaction network. Such nonlinear effects may constitute an inhibition mechanism resulting in asymmetric amplification that is the second key feature of the emergence of homochirality in Frank autocatalysis.

Reactions using alkyl metal reagents came under intense investigation in the 1980's after Mukaiyama[14] (using organomagnesium and organolithium reagents) and Oguni[15] (using organozinc reagents) showed that optically active alcohols could be prepared when chiral amine and amino alcohol additives were included in the reaction mixture. As Wynberg put it, "it does not take a great leap of the imagination"[6] to consider extensions to synthesis of chiral amino alcohols in reactions catalyzed (perhaps self-catalyzed?) by chiral amino alcohols. It is especially worth noting that the second author on the Mukaiyama paper was one Kenso Soai. A decade later, organozinc chemistry had been extensively developed, the concept of nonlinear effects in asymmetric catalysis had been introduced by Kagan,[16] and asymmetric amplification had also been demonstrated in the diethylzinc alkylation of benzaldehyde using chiral amino alcohols. The observation of nonlinear effects leading to asymmetric amplification in catalysis hints at a Frank-type chiral amplification as well as an inhibition mechanism that is key to the model for the emergence of homochirality, which might in fact be considered as the ultimate nonlinear effect. Soai himself had written a 1992 *Chemical Review* on the enantioselective addition of organozinc reagents to aldehydes, which included a section entitled "Asymmetric Self-Catalytic Reaction" in which he noted that if "the structures of the product and the chiral catalyst are the same (chiral self-recatalyst), the reaction system becomes a real self-reproduction system for chiral molecules."[17] Soai had recently introduced the dialkylzinc alkylation of pyridine carbaldehydes (Scheme 6a),[18] which fits criterion 1 of the Frank model, authentic autocatalysis. However, amplification of enantiomeric excess up to or beyond that of the catalyst/product employed had not yet been achieved. Several other substrate classes were also shown to afford autocatalytic reactions (Scheme 6b[19] and 6c[20]), again, however, achieving lower *ee* in the newly formed product than that of the catalyst.

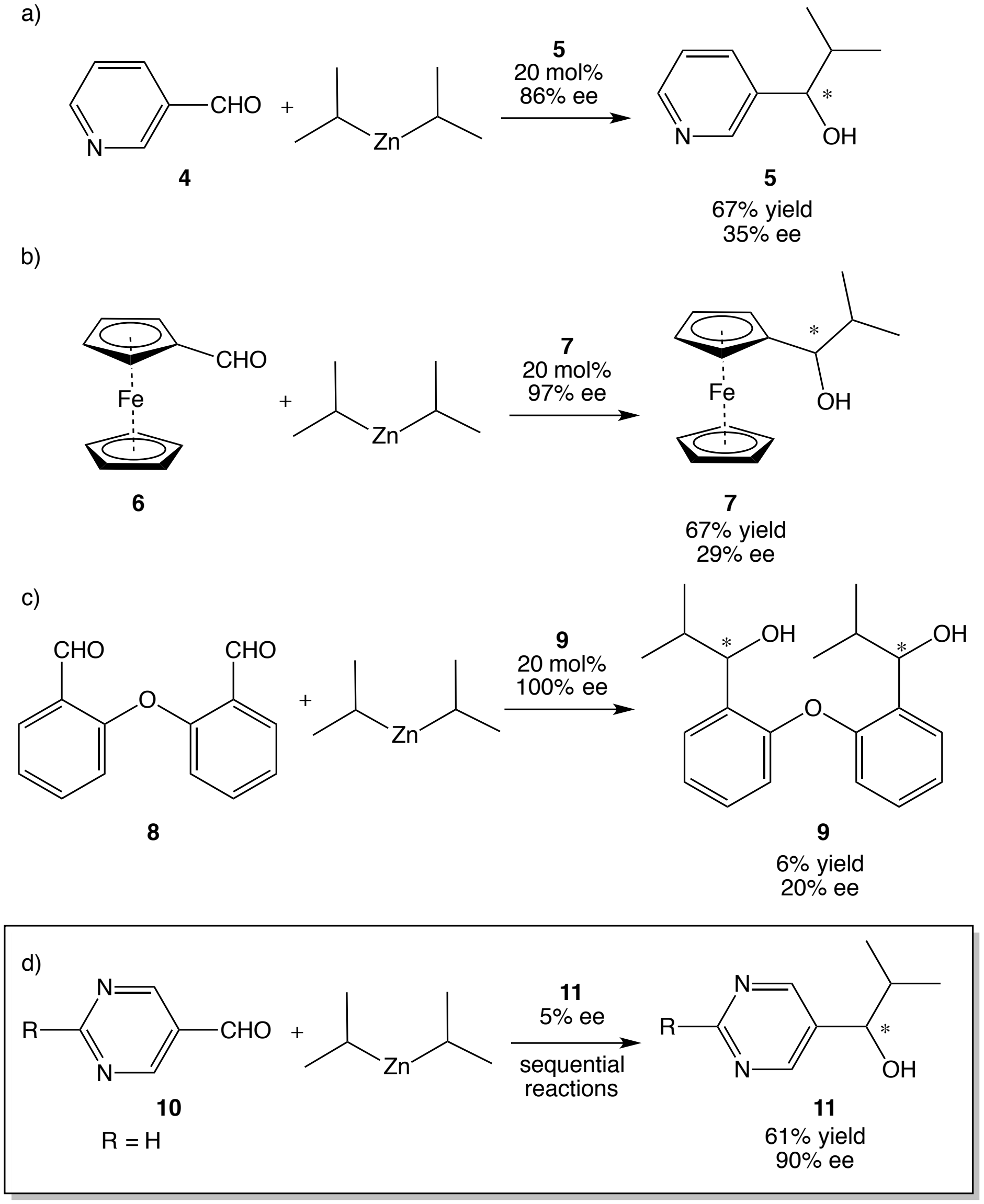

**Scheme 6.** Asymmetric autocatalytic dialkylzinc alkylation reactions reported by Soai. a) pyridine carbaldehydes;[18] b) ferrocenyl aldehydes;[19] c) dialdehydes;[20] d) pyrimidyl aldehydes.[3] The reaction highlighted in part d) is the first example of amplification of product *ee* as well autocatalysis.

Frank's challenge thus went unanswered for 40 years, before it was ultimately met by Kenso Soai with the 1995 *Nature* publication of what has become known as the Soai reaction.[3] The Soai reaction (Scheme 6d), the dialkylzinc alkylation of pyrimidyl aldehydes in which the reaction product catalyzes its own formation, remains the only well-documented example of an experimental system following both criteria 1 and 2 of the Frank autocatalytic model.

Soai's breakthrough came when he extended the substrate scope to include pyrimidyl aldehydes (Scheme 6d).[3] In this case, the reaction rate is accelerated by addition of catalytic amounts of its alcohol product, and the autocatalytic product may be obtained after many cycles in very *high* enantiomeric excess starting from a very *low* enantiomeric excess in the original catalyst. Since this initial discovery, Soai's group has gone on to demonstrate more efficient amplification with other pyrimidyl aldehydes. Higher degrees of asymmetric amplification are found when R=H is substituted with R = $CH_3$ or $C_2$-X, where X = $^{t}$Bu, $C_2Si(CH_3)_3$, or $C_2$(1-adamantyl). However, substrate scope remains limited, and diisopropylzinc remains the only viable alkylating agent leading to asymmetric amplification. Soai has also demonstrated a variety of ways to initiate and direct product enantiomeric excess in this reaction, including exposure to circularly polarized light,[21] by inorganic chiral materials such as quartz,[22] and by the isotope chirality of an initiator molecule.[23] All of these studies reveal that the Soai reaction requires only an extremely small chiral directing force to effect both symmetry breaking and subsequent asymmetric amplification.

This remarkable reaction has been extensively investigated under a wide variety of conditions by a number of groups. Theoretical and modeling studies have been carried out, both in conjunction with experiments and separately. While this reaction has served as an important model for the emergence of homochirality, the fact that its particular chemistry cannot occur in an aqueous prebiotic environment means that the search for more prebiotically relevant reactions with the features of the Frank model is ongoing. In this review, we summarize what we understand about this reaction and its mechanism in order to evaluate why it is such a singular case and to suggest where we might look to uncover a prebiotically plausible version that may align amplifying autocatalysis more closely with studies of prebiotic chemistry and the origin of life.

MECHANISM OF THE SOAI REACTION

The first comprehensive mechanistic rationalization of the intriguing autocatalytic behavior of the Soai reaction came from kinetic, spectroscopic, and computational work by the groups of Blackmond and Brown.[24] The simple autocatalytic kinetic model proposed in this work is an extension of the Kagan $ML_2$ and Noyori reservoir[25] models for nonlinear effects in asymmetric catalysis (Scheme 7). In both models, the formation of dimer or higher order species accounts for the nonlinear relationship between the enantiomeric excess of the added and newly formed catalyst/reaction product (thus addressing the second point in Frank's model, that of inhibition). The main difference between the two models lies in which species serves as the active catalyst. In the Kagan model, monomeric species are neglected, and the higher order species themselves serve as the catalyst, while in the Noyori model, higher order species exist as off-cycle inactive reservoirs that are in equilibrium with on-cycle monomeric catalysts. In either model, the monomer/dimer interactions are characterized by equilibrium constants for homochiral ($K_{homo}$) and heterochiral ($K_{hetero}$) dimer formation, which may in turn be written in terms of an overall dimer constant, $K_{dimer}$, as shown in Scheme 7. The equilibria between monomers and dimers will influence both the concentration and the enantiomeric excess of the active catalyst, whether monomers or dimers. Nonlinearity is possible when unequal partitioning between monomer and dimer species occurs, as dictated by the equilibrium constants.

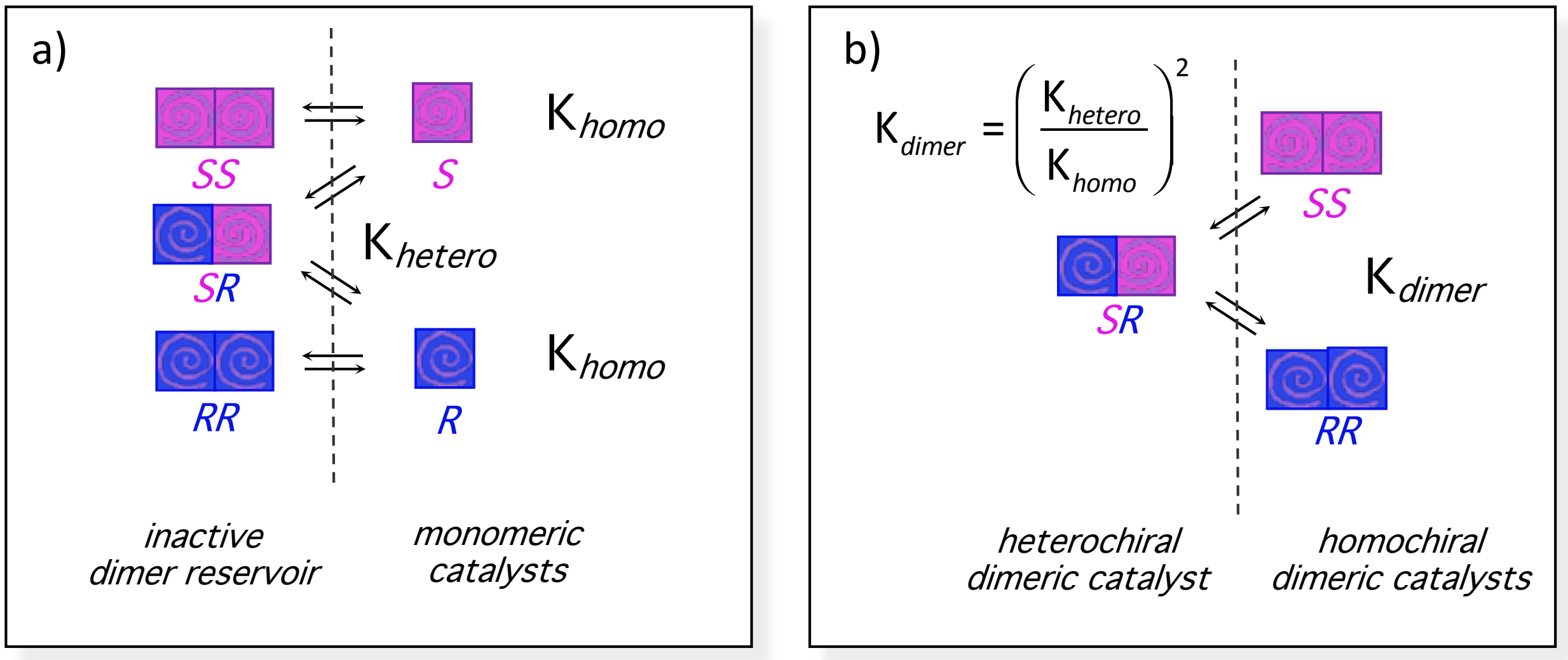

dimer formation:

$$R + R \underset{k_r}{\overset{k_f}{\rightleftarrows}} RR \qquad K_{homo} = \frac{k_f}{k_r}$$

$$S + S \underset{k_r}{\overset{k_f}{\rightleftarrows}} SS \qquad K_{homo} = \frac{k_f}{k_r} \qquad K_{dimer} = \left(\frac{K_{hetero}}{K_{homo}}\right)^2$$

$$R + S \underset{k'_r}{\overset{k'_f}{\rightleftarrows}} SR \qquad K_{hetero} = \frac{k'_f}{k'_r}$$

**Scheme 7.** Models for nonlinear effects in asymmetric catalysis. a) Noyori model,[25] where monomers serve as the active catalysts, in equilibrium with homochiral and heterochiral dimers. b) Kagan model,[16] where monomers are driven to form dimers, which act as the active catalysts. The equilibrium relationships shown dictate the partitioning of *R* and *S* between monomers and dimers. The Kagan model assumes the system is strongly driven towards dimers.

A kinetic model for reaction between pyrimidyl aldehyde (A) and diisopropylzinc (Z) to produce *R* and *S* alkanol products may be written for either monomers or dimers as the active catalysts (Scheme 8). The question of whether it is the monomers or the higher order species that serve as the active catalyst – that is, the Noyori vs. Kagan model – was addressed in the early studies of Blackmond and Brown. Because of the similarity in the chemistry of the Soai reaction to that of the asymmetric alkylation of aldehydes studied by Noyori, a reasonable proposal could be made that the Soai reaction follows a monomer model similar to that proposed by Noyori (Scheme 7a). However, our work provided a number of results that instead provide a compelling case for higher order species themselves rather than the monomers acting as the autocatalysts.

background reaction:

$$A + Z \xrightarrow{k_0} R$$

$$A + Z \xrightarrow{k_0} S$$

autocatalytic reaction (monomer model):

$$A + Z + R \xrightarrow{k_{mon}} R + R$$

$$A + Z + S \xrightarrow{k_{mon}} S + S$$

autocatalytic reaction (dimer model):

$$A + Z + RR \xrightarrow{k_{homo}} R + RR$$

$$A + Z + SS \xrightarrow{k_{homo}} S + SS$$

$$A + Z + SR \xrightarrow{k_{hetero}} \tfrac{1}{2} S + \tfrac{1}{2} R + SR$$

**Scheme 8.** Kinetic model for asymmetric autocatalysis including a racemic background reaction and either monomer- (left) or dimer- (right) catalyzed self-replication.

A first clue pointing to active dimers was the observation that the rate profile for the reaction carried out using enantiopure product as catalyst was found to be almost exactly double that for the racemic catalyst (Figure 1a). We showed that this observation can only occur if a *stochastic* formation of heterochiral and homochiral dimers is formed, that is, $K_{dimer}$ = 4. Stochastic dimer formation means that an *R* monomer has no preference for *R* over *S* in forming an *RR* vs an *SR* dimer. If, for example, the heterochiral species exhibited a significantly higher stability than the homochiral species ($K_{dimer}$ >> 4), the total concentration of reaction product sequestered as dimers in the racemic case would not scale directly with the enantiopure case, as was observed experimentally.

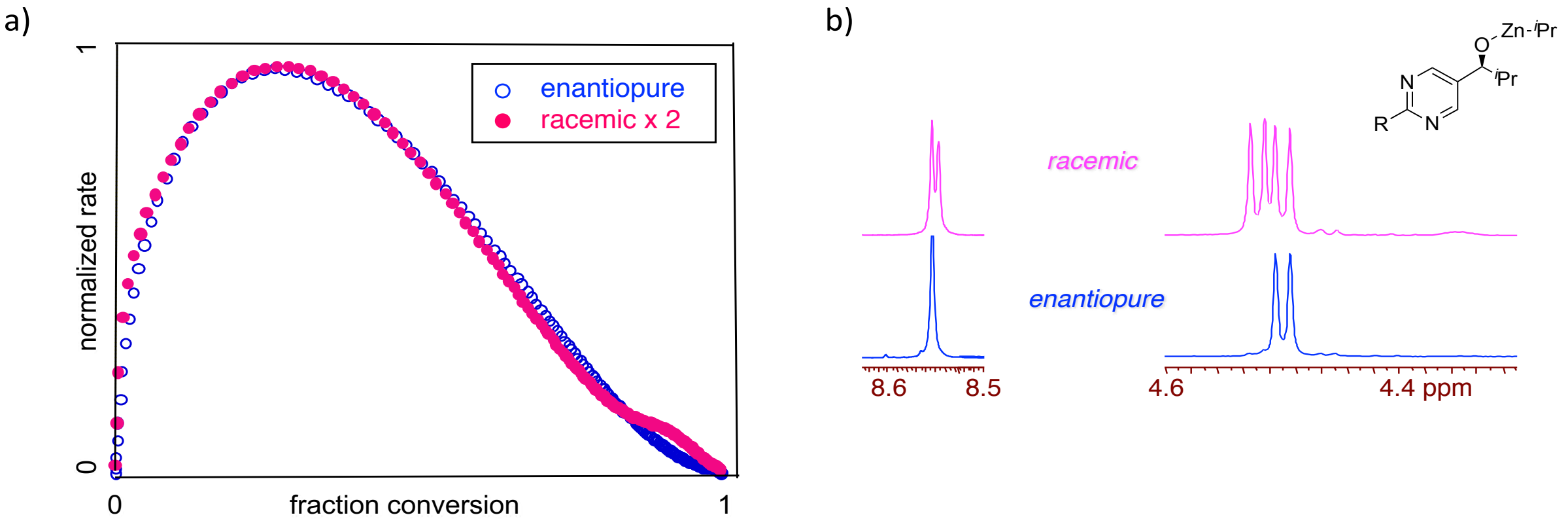

**Figure 1.** Evidence for the stochastic formation of dimers of the Soai reaction product from. Pyrimidyl aldehyde from kinetic[24] (**12**, R=$CH_3$) and spectroscopic[26] (**13**, R=$C_2$TMS ) data. a) normalized plots of rate vs. fraction conversion shows overlay when the racemic profile is multiplied by a factor of 2; b) $^1$H-NMR of racemic and enantiopure product in the arene and alkoxide regions. The racemic case shows double peaks compared to enantiopure, for *RR*+*SS* and *SR*. Integration gives equal (1:1.08) concentrations for *RR*+*SS* and *SR.*

A second clue supporting dimer catalysts came from considering the stochastic ratio of homochiral to heterochiral dimers that is predicted to be formed in a racemic mixture, which is *RR*:*SS*:*SR* = 1:1:2., or homochiral:heterochiral = 1:1. This prediction was validated by NMR spectroscopic studies of such a racemic mixture (Figure 1b), which showed two diastereomeric species of ca. equal concentrations (1:1.08), with one peak ascribed to the sum of enantiomers (*RR*+*SS* = 1+1) and one for *SR* = 2.

This kinetic model demonstrated that homochiral dimers are active catalysts and showed how a small initial chiral bias in the product evolves into increased selection for the major enantiomer. Enantioenrichment occurs as a 1:1 ratio of enantiomeric monomer products is siphoned off into inactive heterochiral species. The fact that the heterochiral dimer is inactive as a catalyst ($k_{hetero}$ = 0) provides the "mutual antagonism" inhibition mechanism required by the Frank model. Fitting experimental kinetic profiles to this dimer kinetic model successfully predicted the temporal evolution of *ee*. Critically, the evolution of *ee* could be predicted solely from the autocatalytic rate profiles for the reaction carried out under a variety of conditions, including different initial *ee* values for the autocatalyst, as shown in Figure 2.[27] It is also important to emphasize that independent validation of the model parameters was obtained under conditions separate from those used to determine the kinetic parameters, which is an essential validation of any kinetic model.

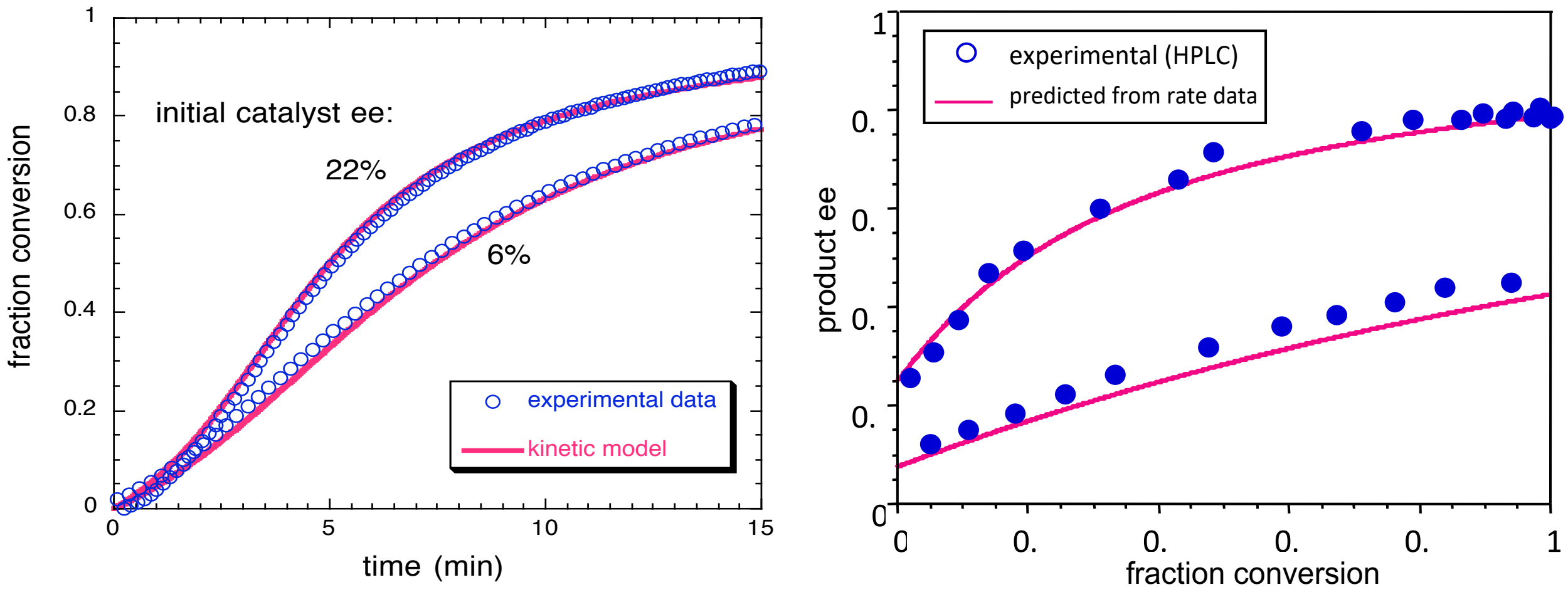

**Figure 2.** Experimental kinetic profiles (symbols) and kinetic model fit (lines) for two different initial autocatalyst ee values in the Soai reaction using 10 mol% **12** with ee values shown as autocatalyst (left).[27] The kinetic model allows prediction of the temporal amplification of ee, which is independently validated by experimental sampling and HPLC analysis (right).

The most compelling argument for the dimer catalyst model, however, comes from the implications of the finding of stochastic dimer formation, which gives $K_{dimer}$ = 4, or $K_{hetero}$ = $2 \cdot K_{homo}$. As mentioned, this equilibrium relation dictates that *R* and *S* will partition equally between the monomer and dimer pools. However, and most critically, a nonlinear effect on product *ee* and the ensuing autocatalytic asymmetric amplification would be able to occur in the monomer model only if partitioning of *R* and *S* species between the pools were found to be *unequal*. Thus, with monomers as catalysts, a system exhibiting stochastic dimer formation ($K_{dimer}$ = 4, or $K_{hetero} = 2 \cdot K_{homo}$) *cannot* produce asymmetric amplification in either a catalytic or an autocatalytic reaction. Thus, while the experimental kinetic data for the Soai reaction could be fit to either a dimer or monomer model, only the dimer model can successfully predict the temporally increasing *ee* that accompanies turnover.[28]

In authentic autocatalysis, the extent of asymmetric amplification possible is limited only by the quantity of reactants that may be fed to the self-replicating system; enantiomeric excess of the Soai reaction product may thus approach homochirality in an asymptotic manner as reaction turnover increases. Ercolani[29] has derived the theoretical relationship shown in Figure 3 between the ultimate product ee, the reaction turnover TON, and the initiating chiral bias $ee_0$ for stochastic formation of active homochiral and inactive heterochiral dimers. Figure 3 demonstrates that the approach to homochirality from a small initial imbalance of enantiomers requires a robust reaction process. Starting from an imbalance of 1 – 0.5% *ee*, the system approaches homochirality after ca. 10,000 turnovers, while an initial imbalance of 0.01% *ee* will reach just over 60% *ee* after the same number of turnovers.

Further mechanistic studies led to the suggestion that dimer catalysts may combine to form tetrameric species[26,30,31,32,33,34] and that in this case the homochiral tetramers may be active as catalysts. Based on substrate concentration dependences, temperature dependence, diffusion coefficients, and calculations, the structure of a tetrameric species was proposed[26,27] (Figure 4) and was later confirmed by crystallization.[35] The tetramer is composed of two dimer structures that had each previously been proposed, a macrocyclic dimer surrounded by two square dimers. Continued oligomerization beyond tetramers was ruled out by studying diffusion behavior of the species formed during reaction.[32] Gridnev has carried out detailed computational studies probing the possible structures of species present under Soai reaction conditions that have shed light on the nature of the active catalyst as well as helped to rationalize the inertness of heterochiral species.[33]

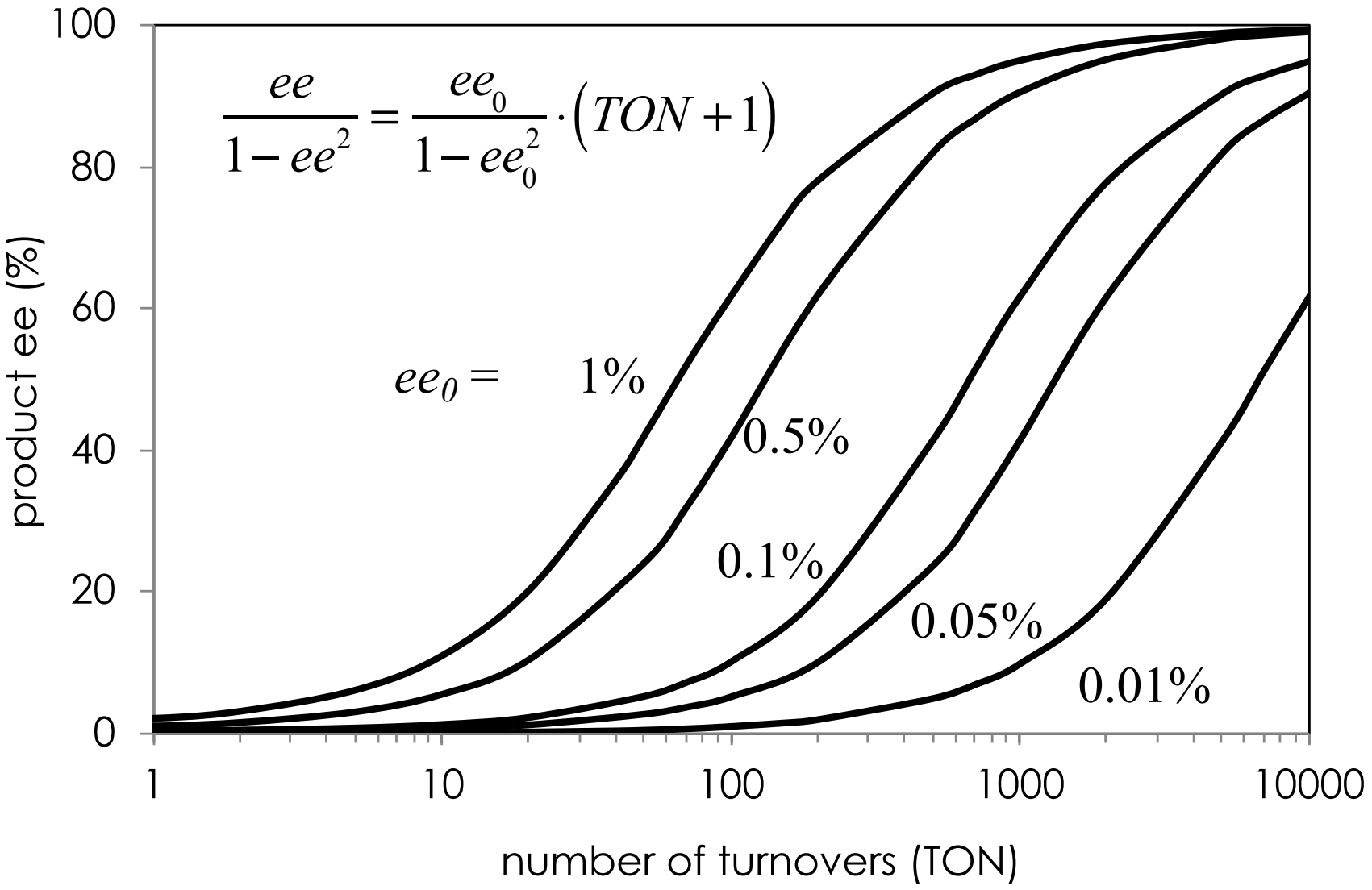

**Figure 3.** Theoretical relationship between product *ee* (y-axis), initial chiral bias $ee_0$ (listed beside each curve) and turnover number TON (x-axis) for the Blackmond/Brown stochastic dimer model of the Soai reaction.[28]

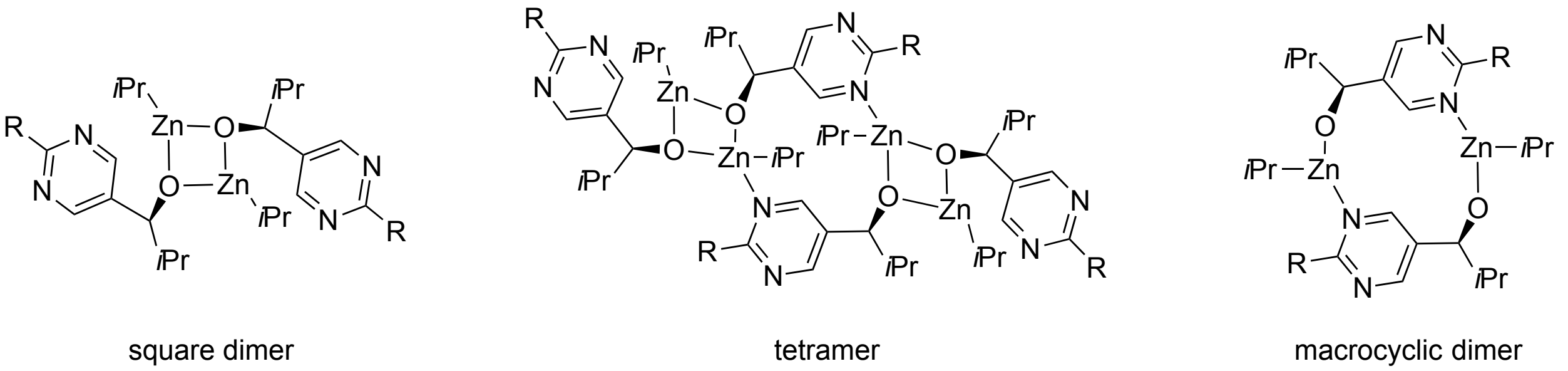

**Figure 4.** Proposed structures formed in the Soai reaction.[26,27,33,35] left: square dimer; center: tetramer; right: macrocyclic dimer.

For autocatalyst **12** with R=$CH_3$, stochastic formation of tetrameric species was predicted, providing the same mathematical description as was found for dimers as catalysts. For bulkier substrates, observation of an increase in the extent of asymmetric amplification beyond that predicted in Figure 3 suggests that equal partitioning of enantiomers between monomeric and higher order species may no longer hold in these cases. Figure 5 compares the stochastic model prediction for *ee* with that found experimentally by Gehring[36] using pyrimidyl catalyst **14** containing the bulky adamantyl group. Amplification is stronger than predicted by the dimer model up to 100 turnovers, after which a decrease in *ee* is attributed to product decomposition. The higher chiral amplification may be attributed to higher stability of the heterochiral species compared to homochiral species for pyrimidyl aldehyde systems with bulkier *para* substitution.

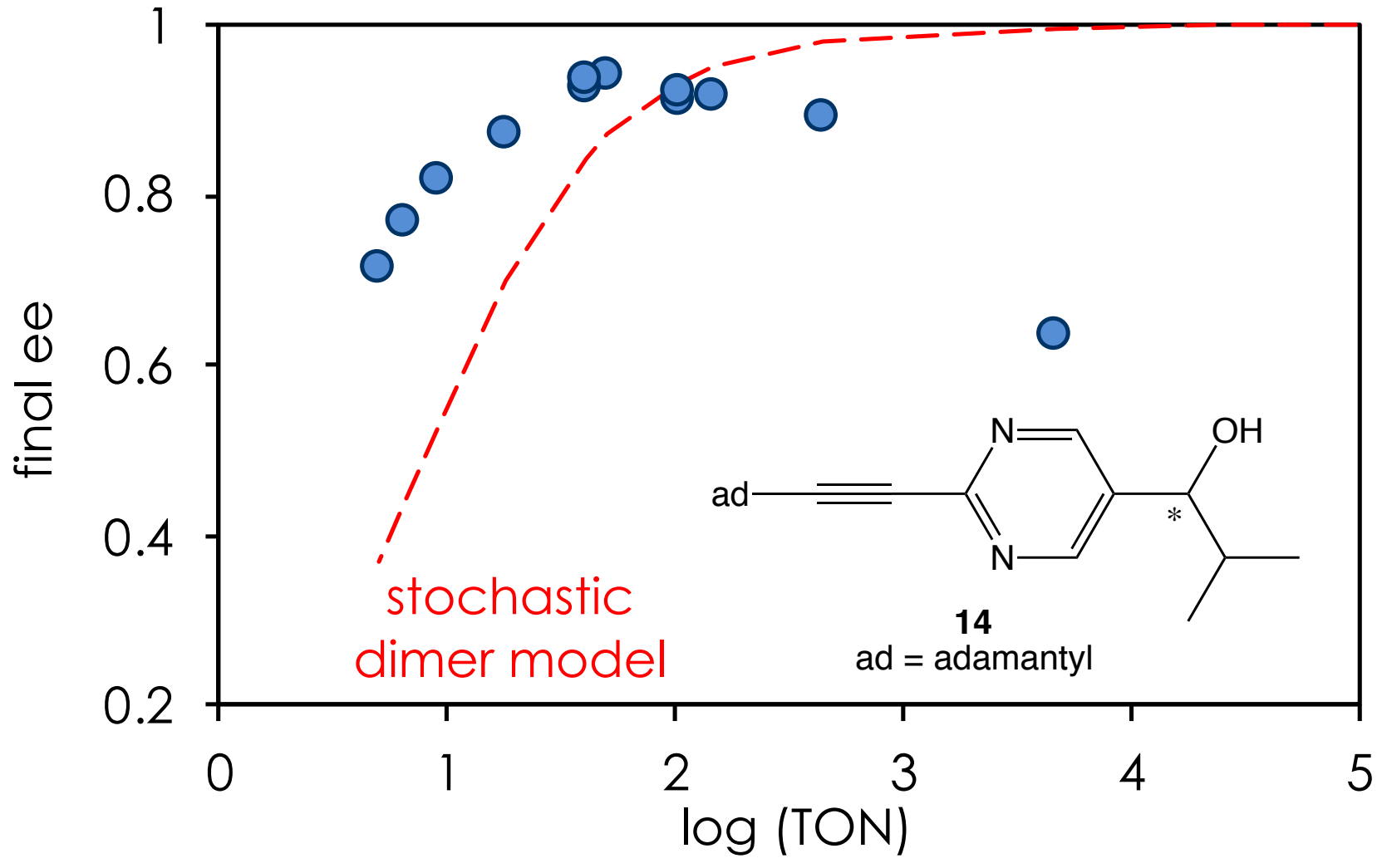

**Figure 5.** Enantiomeric excess as a function of turnover for the Soai reaction using autocatalyst **14** at 7.2% initial enantiomeric excess (blue symbols)[36] compared to that predicted by the stochastic dimer model (calculated here).

Support for the tetrameric model as well as detailed analysis of the origin of enantioselection in the Soai reaction was recently reported by Denmark, Houk and coworkers.[37] As was shown in Scheme 6a, Soai had reported autocatalysis *without* chiral amplification for unsubstituted pyridine carbaldehyde; however, they never revisited this reaction with substituted pyridine carbaldehydes. Interestingly, Denmark found that pyridine-based carbaldehyde substrate **15** with bulky substitution *para* to the aldehyde group not only acted as a tetrameric autocatalyst but also exhibited amplification of *ee* beyond that predicted by the stochastic model (Figure 6). This work helps to demonstrate how *para* substitution increases enantioselection beyond that predicted by the stochastic model. Calculations found that in this case the heterochiral tetramer is more stable by 2.1 kcal/mol, and it is inactive as a catalyst. This work provides a detailed interpretation of how three phenomena – autocatalysis, nonlinear effects, and enantioselection – combine to rationalize the emergence of homochirality in the Soai autocatalytic reaction. They suggest that their analysis may serve as a platform for further studies and explorations in the asymmetric autocatalysis.

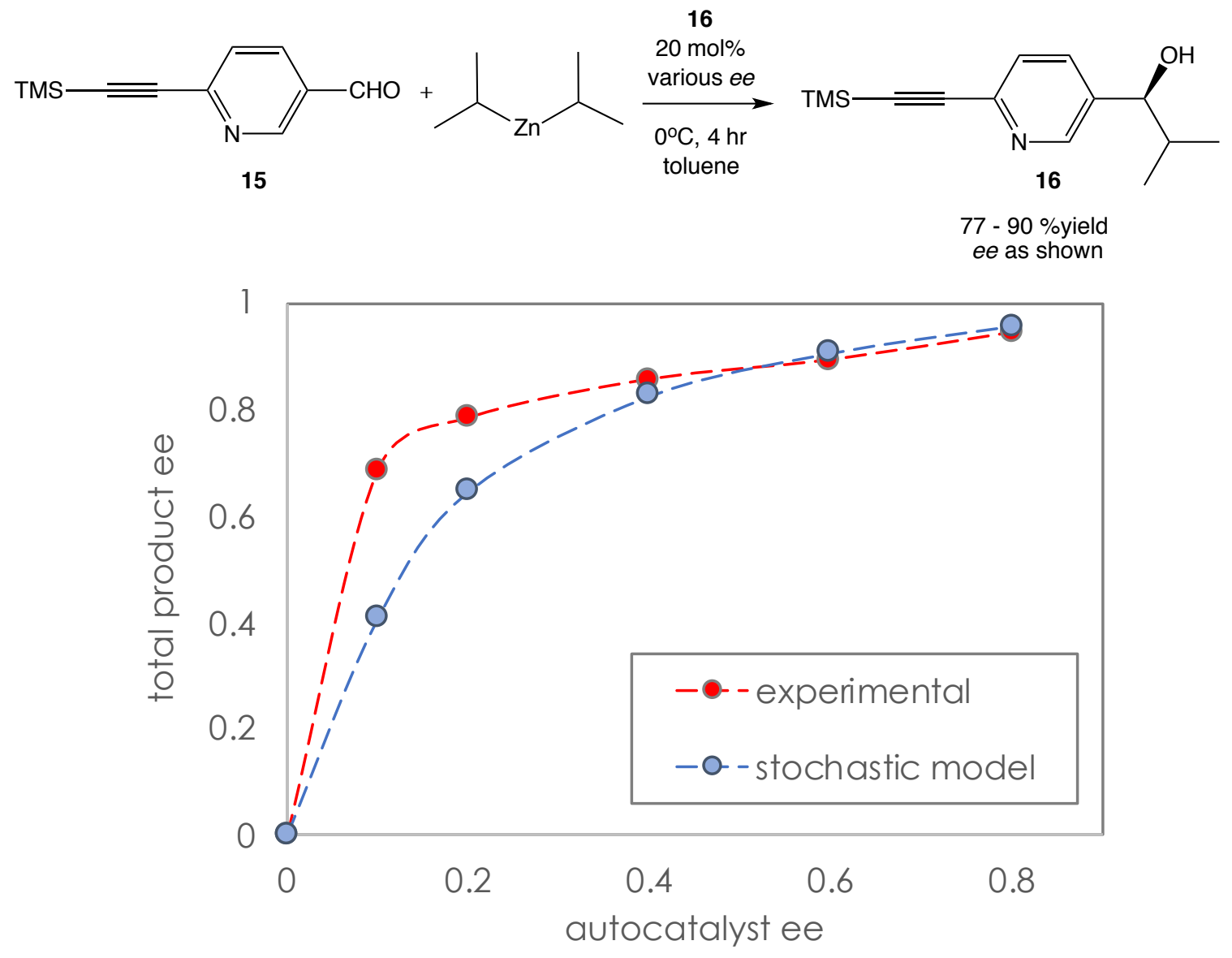

**Figure 6.** Asymmetric amplification in autocatalytic reaction of substituted pyridine carbaldehyde **15**. Experimental result (total product *ee*) taken from Ref. 37 compared to stochastic model prediction of Figure 3 (calculated here).

## OTHER PROPOSED KINETIC MODELS

The main features of the Blackmond/Brown dimer-active kinetic model and its later refinements describing the Soai reaction may be summarized as : 1) the system is driven to form dimer or tetramer species; 2) homochiral dimers/tetramers are active as autocatalysts; 3) heterochiral dimers/tetramers serve as inactive reservoirs; 4) for pyrimidyl substrate R = $CH_3$, formation of dimers and tetramers is stochastic: that is, there is equal preference for *R* and *S* in construction of higher order species; 5) for bulkier R-groups, inactive heterochiral species exhibit higher stability than active homochiral species, leading to more efficient asymmetric amplification.

While this reaction mechanism is supported by voluminous experimental kinetic and spectroscopic data as well as a variety of theoretical calculations, other models have been proposed, including monomer—active or monomer—dominant models. In these cases, however, it may be shown that the proposed model is neither consistent with, nor predictive of, the experimental findings, or that mathematical errors render the model invalid. For example, a model invoking a dimer catalytic species was proposed[38] in which "the concentration of dimers is vanishingly small compared to that of the monomers," in contrast to the experimental spectroscopic findings that higher order (dimer/tetramer) species predominate. A "second-order" kinetic model was developed as shown in Scheme 9, describing the rate of change of the quantity ([R] +[S]), the sum of enantiomeric alkanol products, highlighted in blue. In order to arrive at an analytical solution of this equation, the authors made a simplification equating the quantity ($[R]^2 + [S]^2$) with ($([R]+[S])^2$). As is shown in Scheme 9, however, this assumption sets the term ($2 \cdot [R] \cdot [S]$) equal to zero, which is true only when the system in enantiopure (either [*R*] or [*S*] equals zero).

This assumption is clearly not valid in a model attempting to predict the evolution from low *ee* values to homochirality. The origin of this flawed assumption appears to be confusion between the mathematical operation ($2 \cdot P_R \cdot P_S$), with $P_R$ and $P_S$ designated as the product monomers *R* and *S*, and the variable name chosen for the heterochiral dimer, which was termed $P_R \cdot P_S$ (where "·" signifies a chemical connection between the two enantiomers rather than mathematical multiplication). Thus, the rate equation was developed under the incorrect reasoning that a "vanishingly small" heterochiral dimer concentration would lead to ($2 \cdot P_R \cdot P_S$) ~ 0. This clearly incorrect assumption renders the kinetic model invalid. Thus the proposal that monomers dominate over dimers is supported neither by the experimental data nor the kinetic model.[39,40]

the kinetic model rate equation:

$$\frac{d([R]+[S])}{dt} = k'(a-([R]+[S]))(b-([R]+[S]))([R]^2+[S]^2)$$

...is approximated by:

$$\frac{d([R]+[S])}{dt} \approx k''(a-([R]+[S]))(b-([R]+[S]))([R]+[S])^2$$

....using this equivalence:

$$([R]+[S])^2 = ([R]^2+2\cdot[R]\cdot[S]+[S]^2) \quad \approx \quad ([R]^2+[S]^2)$$

...which is an invalid assumption, except when the system is enantiopure:

$$2\cdot[R]\cdot[S]=0 \quad \Rightarrow \quad [R]=0 \quad or \quad [S]=0$$

**Scheme 9.** Rate equations and underlying assumption made in the kinetic model proposed in Ref. 38. Confusion between a variable name for the heterochiral dimer ($P_R \cdot P_S$) and the mathematical operation ($P_R \cdot P_S$) multiplying *R* and *S* monomers led to a fatal mathematical flaw in the model.

Another model[41] invoking monomers as the active catalyst was developed by authors who did not carry out experimental work but built a mechanistic model based solely on a fit to a single kinetic profile taken from an NMR study by Brown and coworkers.[26] Brown was investigating symmetry breaking in the absence of added product as catalyst. The background reaction forms product that goes on to autocatalyze further turnovers. While racemic product is expected in the absence of an apparent chiral source, often non-zero enantiomeric excesses have been observed in such cases, leading to studies aimed at understanding factors that control how the system may be “tipped” to cause symmetry breaking. It is a challenge for any kinetic model to impose accurate conditions for the autocatalyst concentration in the absence of a defined chiral source in this autocatalytic system, because the direction and extent of the initial symmetry breaking is notoriously susceptible to unmeasurable phenomena including the presence of cryptochiral impurities. Nevertheless, the authors of Ref. 41a aimed to model both symmetry breaking and autocatalysis simultaneously, fitting the single data set shown in Figure 7b to the monomer model of Scheme 8, which contains six adjustable parameters. The fitted parameters allowed calculation of the dimerization equilibria constants shown in Scheme 7. The results of their fit are given in Figure 7b.

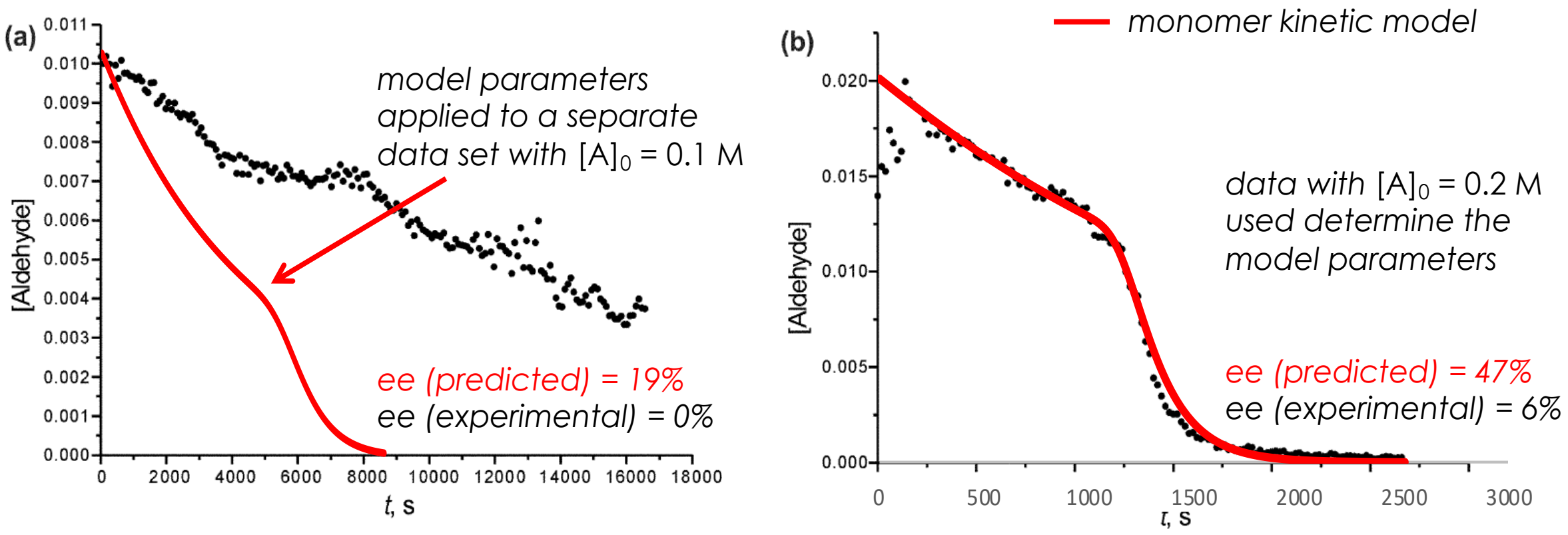

**Figure 7.** Experimental kinetic profiles (black symbols) from Ref. 41a for the Soai reaction carried out in the absence of added reaction product with two different initial concentrations of aldehyde A (R = TMS(alkynyl)) at a) $[A]_0$ = 0.1 M and b) $[A]_0$ = 0.2 M and two equivalents of $^{i}Pr_2Zn$. The data in part b) were used to find kinetic parameters fit to the monomer model of Scheme 8 (red line). The kinetic parameters returned by the model fail to fit the reaction profile in part a) (red line) and these parameters fail to give an accurate prediction of the experimentally measured product *ee* in either case.

This model raises a number of serious concerns because of its glaring inconsistencies with experimental data. While the kinetic model was built entirely from fitting to the single data set shown in part b) of Figure 7, the same set of kinetic rate constants clearly fails to describe a second data set from the same publication, where the identical reaction was carried out using different initial substrate concentrations. It must be emphasized that for a model to have chemical meaning, the same elementary step rate constants must be able to describe data sets from reactions carried out under different conditions, because kinetic constants are not concentration-dependent. It is equally critical to note that this proposed monomer model does not accurately predict the experimentally measured *ee* value found at the end of the reaction under either set of reaction conditions. Further, the kinetic parameters returned a value of $K_{dimer} = 1.4 \times 10^9$, in sharp contrast to the value $K_{dimer} = 4$ found in the Blackmond/Brown dimer kinetic model and experimentally validated by $^1$H-NMR spectroscopic studies as shown in Figure 1.

These authors also stated that they were able to reproduce the 1:1 ratio of heterochiral:homochiral dimers that was found from experimental NMR studies.[41b] This claim is deceptive in that the 1:1 ratio observed spectroscopically in the Blackmond/Brown model was for the case of *racemic* product mixtures, while the monomer model simulation of Ref. 41b provides close to a 1:1 ratio only at *ee* values of greater than 50% *ee* (Figure 8). The relevant comparison of the two models is between racemic systems, and under these conditions, this ratio approaches 20,000:1 in the monomer-active model proposed in Ref. 41, as shown in Figure 8, which is four orders of magnitude greater than the experimentally confirmed value. If this monomer-active model were indeed operative, one would expect to observe a single $^1$H-NMR peak, that for the heterochiral species, which is predicted to dominate under racemic conditions, instead of the two nearly equal peaks experimentally observed, as was shown in Figure 1. From Figure 8 we can conclude that the statement in Ref. 41b that the monomer-active model accurately simulates the experimental NMR observation is categorically incorrect.

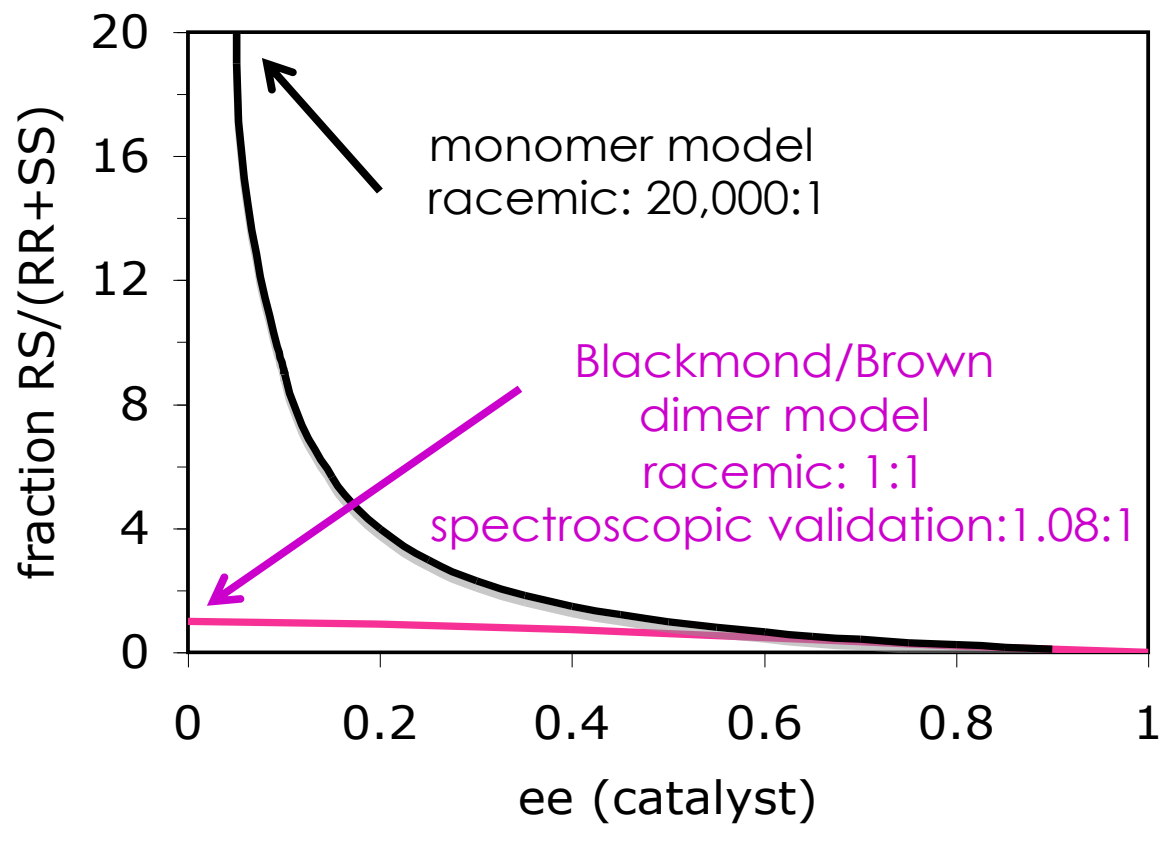

**Figure 8.** Comparison of the ratio of heterochiral to homochiral dimer species in the Blackmond/Brown dimer-active model (pink line) to the monomer-active model of Ref. 41 (black line). The ratio of ca. 1:1 was validated by NMR spectroscopy as was shown in Figure 1.

Given these large discrepancies between the results of the purely computational monomer-active model of Ref. 41 and the overwhelming preponderance of the kinetic and spectroscopic experimental data, it is instructive to probe how such glaring inconsistencies could come about. First, the experimental data in the reaction profile of Figure 7b used in their modeling derives from that of a reaction in which no product was added at the outset, which means that the kinetic model attempts to simulate simultaneously both the process of initial symmetry breaking and that of the subsequent chiral amplification. Deterministic kinetic modeling, such as that employed in Ref. 41, is unable to account for symmetry breaking in the absence of an initial imbalance. How, then, can the model results of Ref. 41 be explained? The answer can be found by scrutinizing the values of the kinetic parameters returned by the model and the implications for relative concentrations of species: although monomers are active as catalysts, the model shows that the system is driven overwhelmingly towards formation of the dimer reservoir so that the concentration of monomers is exceedingly low. For the first 450 minutes of the simulation shown in Figure 7b, the total concentration of monomers *R*+*S* was ca. 5 ppm, or less than 0.025% of the total. Even more tellingly, the concentration *difference* between *R* and *S* monomers, which ultimately determines *ee*, was one in ten million. Such a small absolute difference in concentrations causes instability in the simulation because this difference approaches the level of round-off error of the computer used to carry out the simulation. Thus the force for symmetry breaking used in the computer model of the lab-based experimental reaction data arose from computational fluctuations in concentrations due to computer round-off error. While such an approach may be a useful way to gain theoretical insight into general features caused by instabilities and fluctuations, it is difficult to make a case, as the authors do,[37b] that the particular features of a particular computer modeling program can offer chemical insight into the experimental reaction under study in the laboratory.

In defending the failure of their model parameters to accurately predict the kinetic profile from other experiments such as that shown in Figure 7a, the authors of Ref. 41b maintain that "the data reproduction of multiple experiments with the same rate parameters appeared impossible because of the experimental irreproducibilities", thus suggesting that the computer model offers a more accurate representation of the experimental data than do the data themselves. It is indeed true that symmetry breaking experiments of the Soai reaction in the absence of a chiral source can result in different profiles and different *ee* values for seemingly identical experiments, and such fluctuations are difficult to model and predict. What is unclear from the reasoning of these authors, however, is the question of how they knew to choose the profile of Figure 7b for modeling rather than that of 7a, which would have returned an entirely different set of kinetic parameters and possibly entirely different conclusions: which profile is the one that "correctly" models these fluctuations, and which one is due to what they term "experimental irreproducibilities"? Nevertheless, the authors stated, incorrectly, that "simple autocatalysis involving monomers as the catalytic species is consistent with all reported experimental effects of the Soai reaction." In fact, the only "experimental effect" in agreement with this monomer model is the single kinetic profile that was enlisted to construct the model. If chemical meaning is to be associated with the parameters of a multi-parameter, elementary step reaction model, the model should draw upon multiple, independent data sets, not simply a single experiment. The model must be predictive of the behavior under reaction conditions separate from those used to construct the model. Under these criteria, the monomer-active model of Ref. 41 fails to predict either rate or *ee* under any of the conditions tested.

Such grave errors may be avoided if a more rigorous approach to kinetic modeling of this reaction system is taken. A best practice is to separate attempts to model the symmetry-breaking process, which can be sensitive to undetectable and uncontrollable influences, from modeling of the elementary steps involving catalyst turnover, which can monitored under controlled conditions through accurate measurements of the concentration of product added to the reaction. Both the monomer-dimer equilibria and the product formation rates may then be assessed by deterministic modeling of multiple data sets of controlled reaction conditions and then be validated by using the model parameters to predict the results of further sets of conditions. As was shown in Figure 2, the accuracy of the temporal prediction of *ee* from the parameters of the Blackmond/Brown dimer-active kinetic model lends strong support to the stochastic dimer (or tetramer)-active model. Once validated for well-defined conditions, such a kinetic model may be used in studies aimed at understanding the symmetry breaking process.

## SYMMETRY BREAKING

Probing the nature of the symmetry breaking process has been and remains of great fundamental interest.[26,42,43,44] Deviations from racemic product *ee* are often observed in the Soai reaction carried out in the absence of added product, in most cases giving a stochastic or near stochastic distribution of *ee* values. A key point is to characterize the threshold chiral influence initiating the reaction above which the reaction becomes directed with fidelity toward one hand or the other of the reaction product. The Soai reaction can be initiated from a variety of physical and chemical chiral sources. Perhaps most striking was Soai's studies showing that molecules exhibiting chirality by virtue only of isotopes of the same atom (e.g., H/D, $^{12}C/^{13}C$, $^{14}N/^{15}N$, $^{16}O/^{18}O$) can provide sufficient chiral influence to initiate and dictate the stereochemical outcome of the reaction. In further investigation[45] of the role of these isotopically chiral initiators, we demonstrated that in the initial stages of the reaction, these initiators in fact *inhibit* the autocatalytic pathway by complexing with reaction product formed in the uncatalyzed background reaction. Once the initiator is consumed by saturating as a 2:1 product:initiator complex, subsequently formed reaction product can act as an autocatalyst. A slight difference in stability of diastereomeric product initiator complexes leads to a slight imbalance in the free product enantiomeric excess, which is then amplified in further autocatalytic cycles, as depicted in Scheme 10. The fact that the Soai reaction can be directed with fidelity by barely measurable chiral sources demonstrates that the identity of the directing force may is less critical than is the efficiency of the reaction product in both catalyzing its own formation and in suppressing formation of its opposite enantiomer.

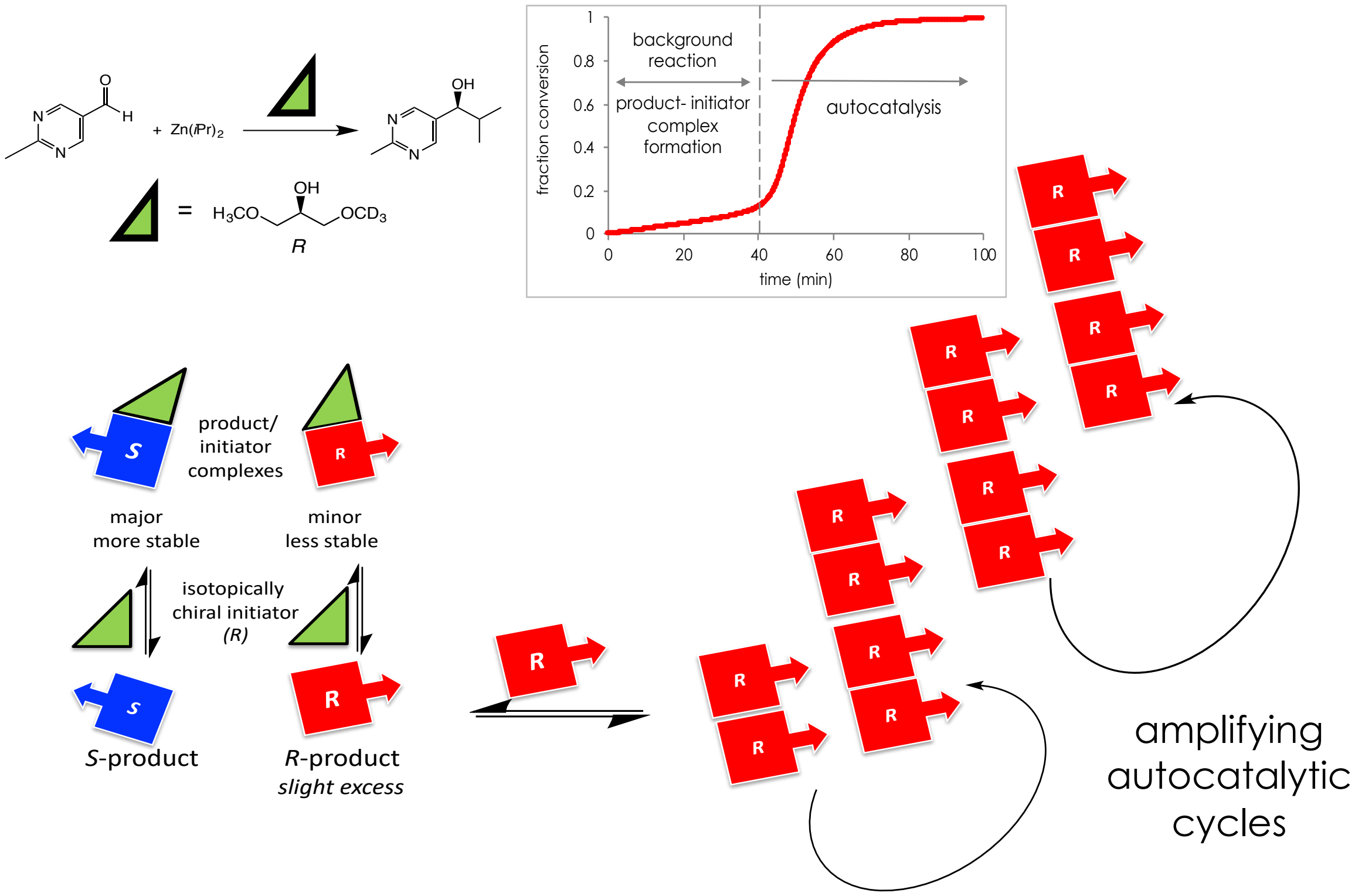

**Scheme 10.** Simplified mechanism of asymmetric amplification in autocatalysis in the presence of chiral initiator (green triangles) that forms complexes with reaction product (blue and red block arrows) produced in the background reaction. Slightly higher stability of the complex formed between the *R* initiator and the *S* product results in a slight excess in free *R* product, which then becomes amplified in the ensuing autocatalytic cycles Reactants not shown for clarity. Inset graph shows an experimental kinetic profile from *in-situ* monitoring of the Soai reaction.

In further studies, we were able to use reactions employing the isotopically chiral initiator at lower and lower ee values to probe the threshold imbalance of product enantiomers that is required to ensure that the chiral amplification proceeds via the stereochemical direction dictated by this imbalance rather than stochastically. Together with stochastic kinetic modeling incorporating the role of stochastic noise, we determined that the threshold autocatalyst *ee* value to escape stochastic behavior lies between 3.5 x $10^{-7}$ – 3.5 x $10^{-8}$ %*ee*.[46] Soai had reported[47] that the reaction began to show inconsistent results when the reaction product was employed at values below $10^{-5}$ %*ee*, presumably hampered by either inevitable inaccuracies in the measurement or the presence of cryptochiral impurities. The fact that Soai's *ee* measurement is two to three orders of magnitude *higher* than the calculated threshold highlights the challenges inherent in accurately and reproducibly carrying out experiments employing such small differences in enantiomer concentrations.

This estimation of the threshold *ee* value required for faithful direction of the Soai reaction also allows us to estimate that the energy required to break symmetry with a consistent chiral bias lies between 1.5 x$10^{-7}$ – 1.5 x $10^{-8}$ kJ/mol. This value may in turn be compared to the best current best estimates for parity-violation energy difference, PVED, caused by asymmetry in the weak force, which lies five to seven orders of magnitude *smaller* than our energy estimate.[48,49] This result may appear to cast doubt on whether the initial chiral selection could have arisen from PVED under the conditions of an amplifying autocatalytic reaction. While we cannot discount PVED as the original source of chiral symmetry breaking, our work provides context for any autocatalytic reaction scenario invoking PVED in symmetry breaking. It is likely that on the time and volume scales of feasible laboratory experiments, this minute energy difference will be lost in the stochastic noise of any autocatalytic network exhibiting the kinetic features of the Soai reaction. It is thus highly unlikely that an experimental observation of PVED-induced symmetry breaking will be observed in Soai autocatalysis.

AUTOCATALYSIS AND REVERSIBILITY

Asymmetric amplification via autocatalysis starting from a small enantiomeric excess necessarily requires many turnovers to achieve high enantioenrichment, as demonstrated in Figure 3 for the stochastic dimer-active model. This means that a significant quantity of the “wrong” enantiomer is produced in the early stages of the reaction, when the autocatalyst ee is low. What if there was some way to convert this product back to its starting material and carry out the autocatalytic reaction again, at later stages with the autocatalyst at higher ee? This idea has been explored in a number of modeling studies,[50,51,52,53] although experimental corroboration has not been possible, since the Soai reaction, the only experimentally documented autocatalytic system demonstrating asymmetric amplification, is not reversible under the conditions employed. If the concept that a “second chance” to achieve high enantioenrichment through reversibility seems to be too good to be true, that is because, indeed, *it is*. The kinetic models that have been constructed to demonstrate this have been shown to be flawed because they violate the principle of microscopic reversibility.[54] Scheme 11 summarizes two autocatalytic models that have been constructed in these studies of the effect of reversibility, along with what we term “missing” reactions, highlighted in red, that were not included in these models.[51,52] The autocatalytic step has been simulated either as first or second order in the autocatalytic reaction product. A critical error in these models arises in how the reverse reaction that regenerates the substrate is written. In Model I, both forward autocatalysis and backward substrate regeneration are written as irreversible steps; however, the backward reaction is not written as the reverse of the forward autocatalytic reaction, but instead as a simple uncatalyzed reaction. In Model II, forward autocatalysis is written as irreversible while substrate recycling is written as a reversible uncatalyzed reaction.

---

**Model I:**

| | | |
|---|---|---|
| forward autocatalysis: $A+L+L \xrightarrow{k_{f,auto}} L+L+L$ <br> $A+D+D \xrightarrow{k_{f,auto}} D+D+D$ | missing reverse autocatalytic reactions: $L+L+L \xrightarrow{k_{r,auto}} A+L+L$ <br> $D+D+D \xrightarrow{k_{r,auto}} A+D+D$ | autocatalysis equilibrium constant: $K_{eq,auto} = \frac{k_f}{k_r} = \frac{[L]_{eq}}{[A]_{eq}} = \frac{[D]_{eq}}{[A]_{eq}}$ |
| proposed recycling of substrate $A$: $A \xrightarrow{k'_f} L$ <br> $A \xrightarrow{k'_f} D$ | missing reverse recycling reactions: $L \xrightarrow{k'_r} A$ <br> $D \xrightarrow{k'_r} A$ | recycling equilibrium constant: $K_{eq,recycle} = \frac{k'_f}{k'_r} = \frac{[A]_{eq}}{[L]_{eq}} = \frac{[A]_{eq}}{[D]_{eq}}$ |

**Model II:**

| | | |
|---|---|---|
| proposed forward (irreversible) autocatalysis: $A+L \xrightarrow{k_{f,auto}} L+L$ <br> $A+D \xrightarrow{k_{f,auto}} D+D$ | missing reverse autocatalytic reactions: $L+L \xrightarrow{k_{r,auto}} A+L$ <br> $D+D \xrightarrow{k_{r,auto}} A+D$ | autocatalysis equilibrium constant: $K_{eq,auto} = \frac{k_f}{k_r} = \frac{[L]_{eq}}{[A]_{eq}} = \frac{[D]_{eq}}{[A]_{eq}}$ |
| proposed (reversible) recycling of substrate $A$: $A \underset{k'_r}{\overset{k'_f}{\rightleftarrows}} L$ <br> $A \underset{k'_r}{\overset{k'_f}{\rightleftarrows}} D$ | | recycling equilibrium constant: $K_{eq,recycle} = \frac{k'_f}{k'_r} = \frac{[A]_{eq}}{[L]_{eq}} = \frac{[A]_{eq}}{[D]_{eq}}$ |

**Model I and Model II:**

relationship between the rate constants:

$$\frac{k_{f,auto}}{k_{r,auto}} = \frac{k'_r}{k'_f}$$

---

**Scheme 11.** Proposed models invoking reversibility in asymmetric autocatalytic reactions. Left: reactions proposed in the model; middle: reverse reactions that are "missing", i.e., not considered in the model; right: equilibrium constants for complete reversible reaction network; bottom: relationship between forward and reverse reaction rate constants dictated by chemical thermodynamics for both models, demonstrating that only three of the four rate constants are independent.

All chemical reactions have the possibility to be reversible. We must reserve this possibility by writing these reaction networks as reversible reactions in order that a description of the equilibrium state can be given; even in cases that remain far from equilibrium under any practical conditions, the equilibrium state sets constraints on the values of the rate constants and their relationship to one another. When the “missing” reverse reactions are included in both Models I and II, the equilibrium relationships dictated by these reactions are given as shown on the right side of Scheme 11. These equilibrium relations emphasize that this condition holds under equilibrium by using the subscript “eq” for the concentrations $[A]_{eq}$, $[L]_{eq}$, and $[D]_{eq}$, even if the possibility of achieving this state is low. These equations may then be manipulated to show that the relationship between these rate constants, as set by the equilibrium condition, is given by the equation at bottom of Scheme 11, which contains only rate constants (no concentration variables), and which holds under all conditions, near to or far from equilibrium. This relationship demonstrates that only three of the four rate constants are independent; once the values for the three rate constants are set, the fourth is fixed by this relationship.

By neglecting the “missing” reverse reactions, both Models I and II have arbitrarily set certain reaction rate constant equal to zero, which violates the constraints set by the equilibrium relationship shown in Scheme 11. This is illustrated in the energy diagram in Figure 9, left, for proposed Model II, where the uncatalyzed reaction is permitted to proceed in both directions, while the reverse reaction rate constant for the autocatalytic case is arbitrarily – and

incorrectly – set equal to zero. The equation at the bottom of Scheme 11 clearly reveals that none of the four rate constants may arbitrarily be set to zero.

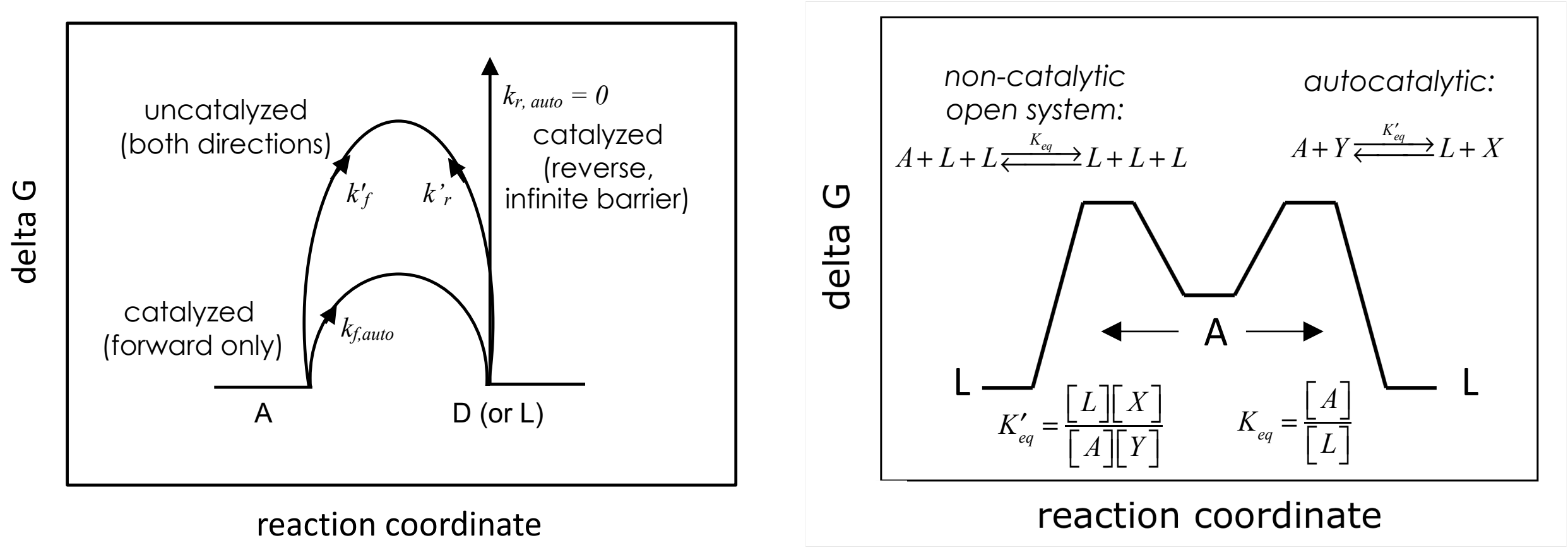

**Figure 9.** Energy diagrams of models for reversible autocatalytic systems. Left: Model II, as proposed in Ref. 53, showing the uncatalyzed reaction as reversible and the autocatalyzed reaction as irreversible, thus violating the principle of microscopic reversibility (redrawn from Ref. 54e). Right, refined model demonstrating how an external chemical driving force can alter the equilibrium relationship for a non-catalytic vs. and autocatalytic reaction, as discussed by Plasson[50] (redrawn from Ref. 54d).

Models that violate tenets of chemical equilibrium such as those shown in Scheme 11 lack experimental realism and cannot provide insight into autocatalytic processes for the emergence of homochirality. A potential "fix" to this conundrum has been suggested by proposing that an external source of energy could drive one but not the other of the reactions shown in the models of Scheme 11.[50,53] Goldenfeld has informally termed this refinement of reversibility models for autocatalytic amplification of *ee* as "pigs *can* fly, with jetpacks,"[53b] introducing an external energy source (the "jetpack") that pushes the autocatalytic system to homochirality under reversible conditions. These authors chose not to define the nature of the external energy source, stating that it "obscures the basic mechanisms leading to

homochirality."[53b] However, this discussion misses a crucial chemical point: only a ***chemical*** source of energy, one that redefines the chemical equilibrium constraint, is a valid energy source to propose as a driver to homochirality in autocatalytic systems such as those proposed in Scheme 11. Chemical energy can be defined as a new reaction that may be designated (at least in a computer model) to act only on the background reaction, or only on the autocatalytic reaction, but not equally on both. Plasson detailed the nature of such a chemical energy source, as shown in Figure 9, right, invoking a system that is open to mass input and output. Substrate A reacts with coupling partner X to produce the enantiomeric product L (or D) and byproduct Y. As shown in Figure 9, this changes the equilibrium relationship for the autocatalytic system so that it is now different from the background reaction. Any non-chemical external energy source would fail to alter the constraint on the rate constants set by the chemical equilibria shown in Scheme 11.

Debenedetti and coworkers[55] have modeled autocatalytic and noncatalytic reaction networks in systems where microscopic reversibility has correctly been taken into account. They demonstrate that strong mutual antagonism as in the Frank model can lead to persistence of asymmetric amplification for geologically relevant time scales even in the presence of reversible reactions, although the final equilibrium state in a closed system will necessarily fluctuate narrowly around the racemic condition.

Goldenfeld points out that any externally driven reaction that is far from equilibrium violates detailed balance and microscopic reversibility and that this is necessary for the emergence of homochirality. Indeed, this is a trivial point: any chemical reaction is necessarily *not* in equilibrium when it is operating under productive conditions. Microscopic reversibility and detailed balance only hold when the system is in equilibrium, but the conditions set by chemical equilibrium – even if this is an aspirational condition – will define the relationships between the rate constants, and those rate constants hold under non-equilibrium conditions. Jet-packs cannot overrule chemical thermodynamics.

OUTLOOK

In the nearly quarter-century since the first report of the Soai reaction, no other system has been documented to exhibit the remarkable features of self-replication with chiral amplification that characterize the Soai reaction. Because the chemistry involved in this reaction has little prebiotic relevance, the search continues for a prebiotically plausible autocatalytic reaction that amplifies enantiomeric excess. The features that such a reaction must exhibit include high selectivity, an inhibition mechanism for production of the minor enantiomer, high efficiency, and persistent self-replication under plausible concentration conditions. The discovery of a reaction or a reaction network embodying these features remains an elusive challenge.

A key to expanding the search for a prebiotically relevant self-replicating system that amplifies enantiomeric excess may lie in the concept of reaction networks rather than in a single, simple autocatalytic reaction.[56] The concept of chemical cycles and their potential role in prebiotic chemistry may be considered as efficient means of producing reservoirs of the organic building blocks required for the increasingly complex chemical transformations that ultimately produced life. Autocatalysis is key to enzymatic chemical cycles with the capacity for informational self-replication. Between these two extreme roles – on one hand the simple mass production and stocking of nutrients, and on the other the complex dynamic process of passing on genetic information – chemical cycles have been bequeathed another role by a group of scientists known as "metabolists," who suggest that self-organizing, interconnected cycles might have developed without the aid, and prior to the evolution, of enzymes.[57] However, equally compelling arguments are made by the opposing camp of "geneticists" that the level of efficiency and specificity demanded for evolutionary success of a small-molecule metabolic cycle has not been demonstrated to date.[58] Indeed, this counterpoint has been seen as another "call to arms"[59] to synthetic organic chemists to participate more directly in the experimentally driven search for chemical reactions and networks that may give rise to the emergence of life.[60]

It is upon this backdrop, focusing on ancient chemistry, that the subject of modern amino acid catalysis introduces the question of how complexity in organic reaction cycles may evolve. Compounds containing simple carbonyl and amine functionalities have been shown to act as both substrates and catalysts. A proline-catalyzed Mannich reaction utilizing acetaldehyde as substrate (Scheme 12a)[61] may be compared with the hydrolysis of diamino maleonitrile, the HCN-tetramer (Scheme 12b).[62] This process starts with the same step as the Mannich reaction, but the roles of catalyst and substrate are reversed in the cycle that ensues. The hydrolysis of diamino maleonitrile is arguably an important step for a putative synthesis of RNA.

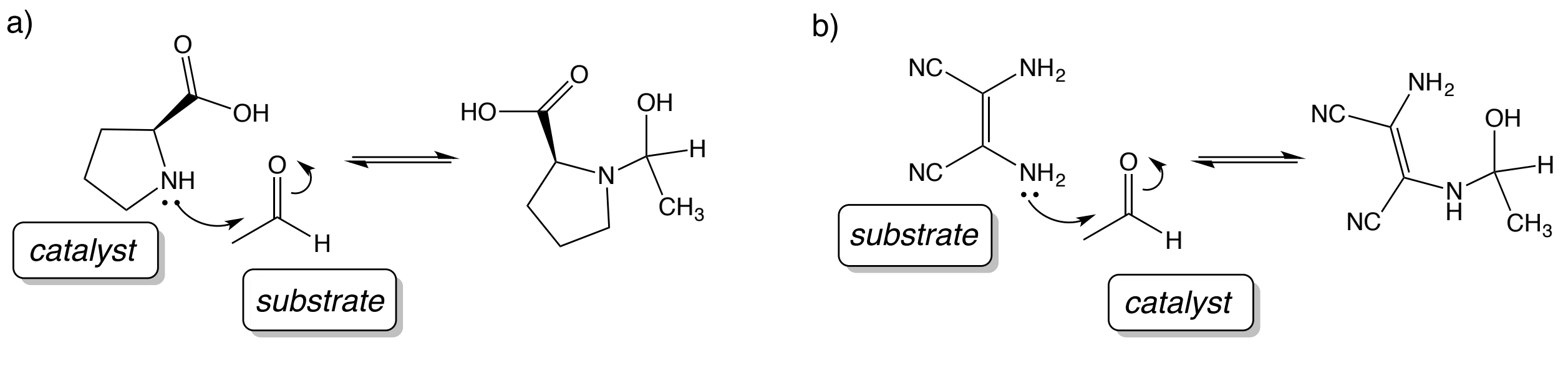

**Scheme 12.** The roles of amine and carbonyl compounds in asymmetric reactions. a) amine as catalyst, carbonyl as substrate ; and b) the reverse.

While these studies have not directly addressed chirality, they explore the possible roots of prebiotic self-organizing chemical processes. Indeed, this chemistry takes us back to the very first organic catalytic reaction, reported in 1860 by Justus von Liebig, where acetaldehyde was used as a catalyst to convert cyanogen and water to the diamide of oxalic acid.[63] The Strecker reaction of 1850 proceeds by a similar mechanism to produce amino acids from aldehydes and ammonia.[64] Thus an important chemical platform proceeding through aldehyde-amine hemiaminal intermediates exhibits a dual role as an erstwhile nutrient (reactant) and enabler (catalyst), offering food for thought concerning the emergence of viable autocatalytic cycles as coupled or cascade networks (Scheme 13). Introducing asymmetry into these systems might expand a Frank-type model to incorporate a coupled series of autocatalytic reactions from which homochirality could emerge.

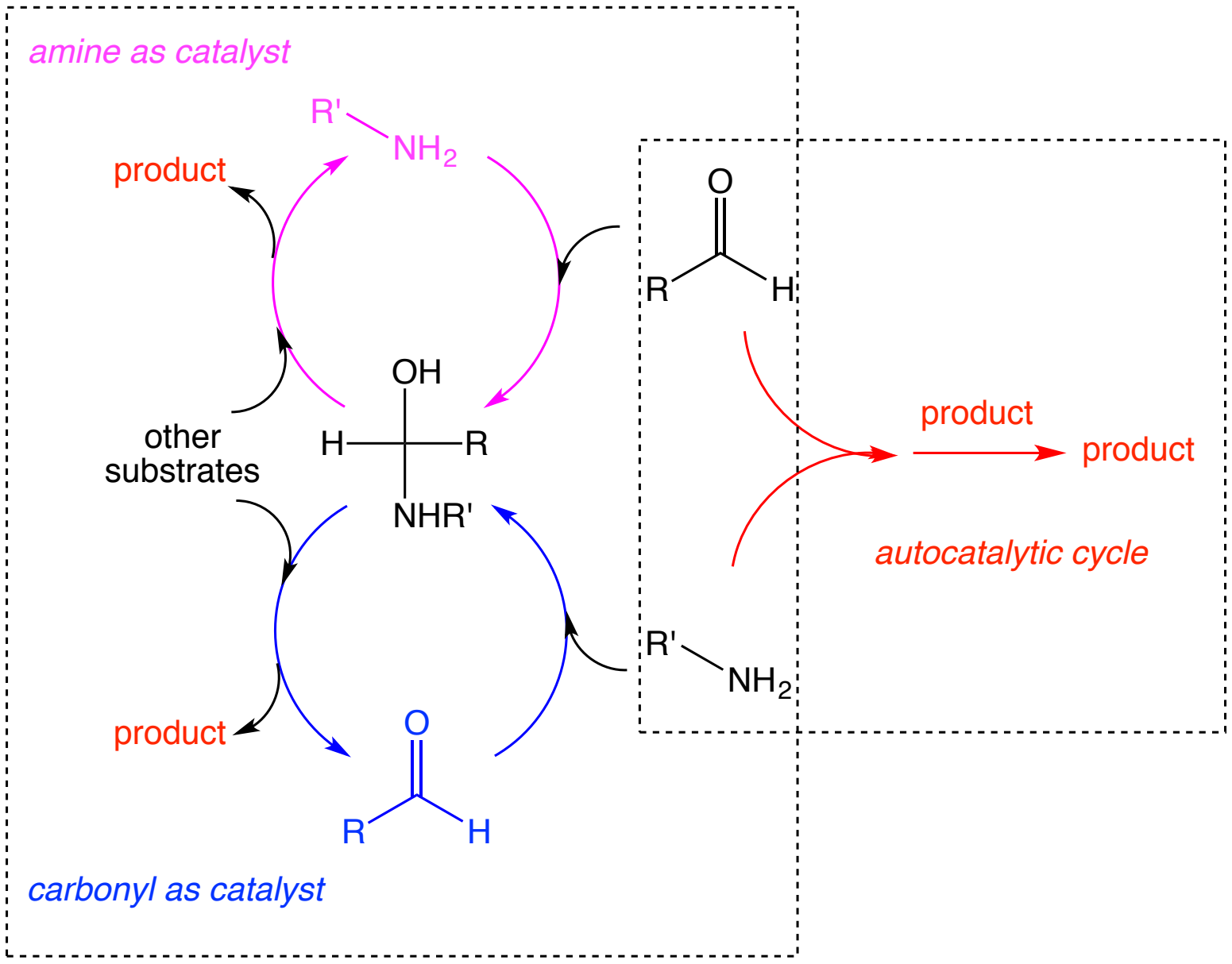

**Scheme 13.** Proposed cycles with carbonyl and amine compounds serving alternatively as catalysts and substrates in a coupled cascade, with the possibility of autocatalytic self-replication and the amplification of enantiomeric excess.

SUMMARY

Autocatalysis with asymmetric amplification provides an elegant solution to the question of the emergence of homochirality characterized by two features: i) a symmetry breaking transition that is highly sensitive to small asymmetric influences; and ii) a reaction that exhibits a higher order burst of autocatalytic activity. The first feature allows for high selectivity; the second is required to maintain this selectivity above stochastic noise and to propagate the selective pathway at the expense of that of its enantiomer. While the Soai reaction remains the sole documented example of the theoretical Frank model for spontaneous asymmetric synthesis, the search to discover an autocatalytic network that exhibits these features and at the same time exemplifies prebiotically plausible chemistry is ongoing. Homochirality might have involved not a single reaction or event but rather a series of persistent chemical and physical processes acting synergistically as a key part of the emergence of life on earth.

ACKNOWLEDGMENTS

Funding from the Simons Foundation Collaboration on the Origins of Life (SCOL Award 287625) is gratefully acknowledged. Collaboration and stimulating discussions with J.M. Brown, A. Armstrong, O.K. Matar, D.L.Kondepudi, R. Breslow, and D. Hochberg and acknowledged.

[3] Kondepudi, D.K., Kaufman, R.J., Singh, N. 1990. Chiral symmetry breaking in sodium chlorate crystallization. *Science* **250:** 975-76.

[4] Soai, K. Shibata, T., Morioka H., Choji, K. 1995. Asymmetric autocatalysis and amplification of enantiomeric excess of a chiral molecule. *Nature*, **378:** 767-68.

[5] Ball, P. Giving life a hand. 2007 (March), *Chem. World*, 30–31.

[6] Wynberg, A. 1989. Asymmetric autocatalysis: facts and fancy. J. *Macromolecular Science, Part A,* **A26(8):** 1033-1041.

[7] a) Mower, M., Blackmond, D.G. 2015. Mechanistic rationalization of unusual sigmoidal kinetic profiles in the Machetti-de Sarlo cycloaddition reaction. *J. Am. Chem. Soc.,* **137:** 2386-2391b: Blackmond D.G. 2009. An Examination of the Role of Autocatalytic Cycles in the Chemistry of Proposed Primordial Reactions. *Angew. Chemie. Int. Ed.* **48:** 386-390; c) Bissette, A.J., Fletcher, S.P. 2013. Mechanisms of autocatalysis. *Angew. Chemie. Int. Ed.* **52:** 12800-12826.

[8] a) Rosner, T.; Pfaltz, A.; Blackmond, D.G. 2001. Observation of unusual kinetics in Heck reactions of aryl halides: the role of non-steady-state catalyst concentration. *J. Am. Chem. Soc*. **123:** 4621-4622; b) Yoon, H. J.; Mirkin, C. A. 2008. PCR-like cascade reactions in the context of an allosteric enzyme mimic, *J. Am. Chem. Soc*. **130**: 11590-11591.

[9] a) Iwamura, H. Mathew, S.P., Armstrong, A., Blackmond, D.G. 2004. *In-situ* catalyst improvement in the proline-mediated $\alpha$-amination of aldehydes. *J. Am. Chem. Soc*. **126:** 11770-11771; b) Mathew, S.P. Klussmann, M,. Iwamura, H., Wells, D.H., Jr., Armstrong, A., Blackmond, D.G. 2006. A mechanistic rationalization of unusual kinetic behavior in proline-mediated C-O and C-N bond-forming reactions. *Chem. Commun.* 4291-4293.

[10] Breslow, R. 1959. On the mechanism of the formose reaction. *Tetrahedron Letters*. **1**: 22–26.

[11] a) Breslow, R.; Cheng, Z.-L. L-Amino acids catalyze the formation of an excess of d-glyceraldehyde, and thus of other D sugars, under credible prebiotic conditions. 2010. *Proc. Natl. Acad. Sci. U.S.A.* **107**: 5723–5725; b) Blackmond, D. G.; Moran, A.; Hughes, M.; Armstrong, A. Unusual reversal of enantioselectivity in the proline-mediated r-amination of aldehydes induced by tertiary amine additives. J*. Am. Chem. Soc*. 2010, **132**: 7598–7599; c) Hein, J.E., Blackmond, D.G., 2012. On the Origin of Single Chirality of Amino Acids and Sugars in Biogenesis. *Acc. Chem. Res*., **45:** 2045-2054.

[12] Betti, M., 1900. On the addition of benzyl amine to naphthol. *Gazz. Chim. Ital.,* **II:** 301.

[13] Pascal, R., Taillades, J., Commeyras, M. 1978. Systemes de strecker et apparentes—X: Decomposition et hydratation en milieu aqueux basique des α-aminonitriles secondaires. Processus d'hydratation autocatalytique et catalyse par l'acetone. *Tetrahedron*, **34**: 2275-2281.

[14] Mukaiyama, T., Soai, K. Sato, T. Shimizu, H. Suzuki, K. 1979. Enantioface-differentiating (asymmetric) addition of alkyllithium and dialkylmagnesium to aldehydes by using (2S,2'S)-2-hydroxymethyl-1-[(1-alkylpyrrolidin-2-yl)methyl]pyrrolidines as chiral ligands, *J. Am. Chem. Soc.*, **101**: 1455-1460.

[15] Oguni, N., Omi, T., Yamamoto, Y., Nakamura, A. 1983. Enantioselective addition of diethylzinc to arylaldehydes catalyzed by chiral cobalt(II) and palladium(II) complexes. *Chem. Lett.,* 841.

[16] (a) Puchot, C.; Samuel, O.; Duñach, E.; Zhao, S.; Agami, C.; Kagan, H. B. 1986. Nonlinear effects in asymmetric synthesis. examples in asymmetric oxidations and aldolization reactions. *J. Am. Chem. Soc*. **108**: 2353−2357. (b) Guillaneux, D.; Zhao, S. H.; Samuel, O.; Rainford, D.; Kagan, H. B. 1994. Nonlinear effects in asymmetric catalysis. *J. Am. Chem. Soc*. **116**: 9430−9439.

[17] Soai, K. Niwa, S. Enantioselective addition of organozinc reagents to aldehydes, 1992. *Chem. Rev.* **92**: 833-856.

[18] Soai, K. Niwa, S. Hori, H. 1990. Asymmetric self-catalytic reaction. self-production of chiral 1-(3-pyridyl)alkanols as chiral self-catalysts in the enantioselective addition of dialkylzinc reagents to pyridine-3-carbaldehyde. *JCS. Chem. Commun*. 682-683.

[19] Soai, K., Hayase, T. Takai, K. 1995. Catalytic chirally self-replicating molecule. asymmetric autocatalytic reaction of a zinc alkoxide of chiral 1-ferrocenyl-2-methylpropan- 1-ol. *Tetrahedron: Asymm.,* **6**: 637-638.

[20] Soai, K., Hayase, T., Shimada, C., Isobe, K. 1994. Catalytic asymmetric synthesis of chiral diol, bis[2-(1-hydroxyalkyl)-phenyl]ether, an asymmetric autocatalytic reaction, *Tetrahedron: Asymm*., **5**: 789-792.

[21] Shibata, T.; Yamamoto, J.; Matsumoto, N.; Yonekubo, S.; Osanai, S.; Soai, K. 1998. Amplification of a slight enantiomeric imbalance in molecules based on asymmetric autocatalysis: the first correlation between high enantiomeric enrichment in a chiral molecule and circularly polarized light. *J. Am. Chem. Soc*. **120**: 12157−12158.

[22] Soai, K.; Osanai, S.; Kadowaki, K.; Yonekubo, S.; Shibata, T.; Sato, I. 1999. d- and l-Quartz-promoted highly enantioselective synthesis of a chiral organic compound. *J. Am. Chem. Soc.* **121**: 11235−11236.

[23] Kawasaki, T.; Matsumura, Y.; Tsutsumi, T.; Suzuki, K.; Ito, M.; Soai, K. 2009. Asymmetric autocatalysis triggered by carbon isotope (13C/12C). *Science* **324:** 492−495.
[24] Blackmond, D.G., McMillan, C. R. Ramdeehul, S. Schorm, A., Brown, J.M. 2001. Origins of asymmetric amplification in autocatalytic alkylzinc additions, *J. Am. Chem. Soc*. **123:** 10103-10104.

[25] Kitamura, M.; Suga, S.; Oka, H.; Noyori, R. 1998. Quantitative analysis of the chiral amplification in the mino-alcohol-promoted asymmetric alkylation of aldehydes with dialkylzincs. J. *Am. Chem. Soc.* **120:** 9800−9809.

[26] Gridnev, I. D.; Serafimov, J. M.; Quiney, H.; Brown, J. M. 2003. Reflections on spontaneous asymmetric synthesis by amplifying autocatalysis. *Org. Biomol. Chem.* **1**: 3811−3819.

[27] Buono, F. G.; Blackmond, D. G. 2003. Kinetic evidence for a tetrameric transition state in the asymmetric alkylation of pyrimidyl aldehydes. *J. Am. Chem. Soc*. **125**: 8978−8979.

[28] a) Blackmond, D. G. Asymmetric autocatalysis and its implications for the origin of homochirality. 2004. *Proc. Natl. Acad. Sci. U. S. A.* **101**: 5732−5736. b) Gridnev, I. D.; Brown, J. M. 2004. Solution structure and reagent binding of the zinc alkoxide catalyst in the Soai asymmetric autocatalytic reaction. *Proc. Natl. Acad. Sci. U. S. A.* **101**: 5727−5731.

[29] Schiaffino, L., Ercolani, G. 2009. Amplification of chirality and enantioselectivity in the asymmetric autocatalytic soai reaction. *ChemPhysChem,* **10**: 2508-2515.

[30] Klankermayer, J.; Gridnev, I. D.; Brown, J. M. 2007. Role of the isopropyl group in asymmetric autocatalytic zinc alkylations. *Chem. Commun.* 3151−3153.

[31] Gridnev, I.D.; Serafimov, J.M.; Brown, J.M. 2004. Solution structure and reagent binding of the zinc alkoxide catalyst in the Soai asymmetric autocatalytic reaction. *Angew. Chem. Int. Ed.,* **43**: 4884–4887.

[32] Quaranta, M.; Gehring, T.; Odell, B.; Brown, J.M.; Blackmond, D.G. 2010. Unusual inverse temperature dependence on reaction rate in the asymmetric autocatalytic alkylation of pyrimidyl aldehydes. *J. Am. Chem. Soc.* **132**: 15104-15107.

[33] Gridnev, I.D.; Vorobiev, A.Kh. 2012. Quantification of sophisticated equilibria in the reaction pool and amplifying catalytic cycle of the Soai reaction. *ACS Catal.,* **2**: 2137-2149.

[34] Gehring, T.; Quaranta, M.; Odell, B.; Blackmond, D. G.; Brown, 2012. J. M. Observation of a transient intermediate in Soai's asymmetric autocatalysis: insights from 1H NMR turnover in real time. *Angew. Chem., Int. Ed.* **51**: 9539−9542.